# Palmitic Acid Sophorolipid Biosurfactant: From Self-Assembled Fibrillar Network (SAFiN) To Hydrogels with Fast Recovery

Niki Baccile,[1,]* Ghazi Ben Messaoud,[1,†] Patrick Le Griel,[1] Nathan Cowieson,[2] Javier Perez,[3] Robin Geys,[4] Marilyn De Graeve,[4] Sophie L. K. W. Roelants,[4,5] Wim Soetaert[4,5]

[1] Sorbonne Université, Centre National de la Recherche Scientifique, Laboratoire de Chimie de la Matière Condensée de Paris, LCMCP, F-75005 Paris, France

[2] Diamond Light Source, Harwell Science and Innovation Campus, Didcot, Oxfordshire, OX11 0DE, UK

[3] Synchrotron Soleil, L'Orme des Merisiers, Saint-Aubin, BP48, 91192 Gif-sur-Yvette Cedex, France

[4] Ghent University, Centre for Industrial Biotechnology and Biocatalysis (InBio.be), Coupure Links 653, Ghent, Oost-Vlaanderen, BE 9000

[5] Bio Base Europe Pilot Plant, Rodenhuizekaai 1, Ghent, Oost-Vlaanderen, BE 9000

*Correspondence to: Dr. Niki Baccile, niki.baccile@sorbonne-universite.fr, Phone: 00 33 1 44 27 56 77

† Current address: DWI- Leibniz Institute for Interactive Materials, Forckenbeckstrasse 50, 52056, Aachen, Germany

**Abstract**

Nanofibers are an interesting phase into which amphiphilic molecules can self-assemble. Described for a large number of synthetic lipids, they were seldom reported for natural lipids like microbial amphiphiles, known as biosurfactants. In this work, we show that the palmitic acid congener of sophorolipids (SLC16:0), one of the most studied families of biosurfactants, spontaneously forms a self-assembled fiber network (SAFiN) at pH below 6 through a pH jump process. pH-resolved *in-situ* Small Angle X-ray Scattering (SAXS) shows a continuous micelle-to-fiber transition, characterized by an enhanced core-shell contrast between pH 9 and pH 7 and micellar fusion into flat membrane between pH 7 and pH 6, approximately. Below pH 6, homogeneous, infinitely long nanofibers form by peeling off the membranes. Eventually, the nanofiber network spontaneously forms a thixotropic hydrogel with fast recovery rates after applying an oscillatory strain amplitude out of the linear viscoelastic regime (LVER): after being submitted to strain amplitudes during 5 min, the hydrogel recovers about 80% and 100% of its initial elastic modulus after, respectively, 20 s and 10 min. Finally, the strength of the hydrogel depends on the medium's final pH, with an elastic modulus fivefold higher at pH 3 than at pH 6.

**Introduction**

Stimuli-responsive peptides, proteins and lipids[1–4] attract a large interest in the field of nanotechnology for their ability to self-assemble into 2D and 3D soft materials, which can in turn be employed in a growing number of high-tech applications,[5] such as protective coating for cells,[6] regenerative medicine,[7] lab-on-a-membrane prototyping[8] or self-healing materials.[9] Lipids are particularly interesting compounds because they are ubiquitous natural molecules, which are still easy to synthesize and, despite their often simple structure, they can self-assemble into a variety of soft architectures,[10] possibly leading to complex isotropic (e.g., disordered entangled fibers) or anisotropic (e.g., lamellar) nano and meso-structures having interesting mechanical properties and potential applications.[11–13]

Nanofibers, whether reported in the shape of nanotubes,[14] twisted or helical ribbons,[15] are an interesting form of self-assembled structures found in lipids,[16,17] peptides,[18] lipid peptides[19] and proteins.[20] Generally driven by both specific (H-bonding, π-π stacking) intermolecular forces and non-specific (steric hindrance, hydrophobic effect) interactions,[12,13,21] sometimes driven by interactions with chiral counterions,[22] self-assembled fibrillar networks (SAFiN) are often involved in the development of soft gelled materials,[12,13,23–25] with applications as scaffolds for tissue engineering,[26] wound healing,[27] cancer treatment.[28]

Microbial glycolipids are a class of compounds generally addressed to as microbial amphiphiles, or biosurfactants. They all have in common their microbial fermentation production source from vegetable oils and glucose.[29,30] Commonly considered as safe compounds from a cytotoxicity point of view,[31,32] their high-end applicative potential is still to be unveiled and one of the main reasons is the poor knowledge of their phase behaviour in water under application-relevant conditions, that is at volume fractions below 10 wt%, at pH between 5 and 8 and ionic strength above the mM range. A set of recent studies has shown the ability of a wide range of microbial glycolipids to form micelles,[33,34] ribbons,[35] vesicles,[36–38] sponge,[39] lamellar[39,40] structures and even simple[41] or complex coacervates.[42]

Fibrillation in the form of nanoribbons and nanohelices from microbial biosurfactants has been shown for cellobioselipids,[36,43] symmetrical sophorolipids,[44] amine-derivatives of sophorolipids[45] and, above all, the neutral form of stearic acid (C18:0) derivative of sophorolipids.[35,46,47] The latter shows fibrillation at both acidic and alkaline pH for, respectively, the –COOH-ending[35,46] and –NH₂-ending[47] congeners. However, hydrogel formation from microbial biosurfactants in the absence of additives (e.g., gelators, polymers) was mentioned for cellobioselipids[48] and more deeply studied for C18:0 glucolipids (lamellar gel),[40,49] symmetrical C16:0[44] and acidic C18:0 sophorolipids (fibrillar gels).[50] In the latter cases, it was shown that the gel strength depends on either the temperature or pH variation rates, in agreement with previous studies on low-molecular weight gelators (LMWG), of which the gel strength was shown to be related to the content of spherulitic structures, which are in turn controlled by the supersaturation extent and, consequently, by kinetics.[51]

In this work, we address the self-assembly properties of the recently discovered non-acetylated palmitic acid sophorolipids, SLC16:0, derived by fermentation of *S. bombicola* CYP1BMR in the presence of palmitic acid and glucose (Figure 1).[52] We employ pH-resolved *in-situ* small angle X-ray scattering (SAXS) to follow its self-assembly in water at room temperature. We find that nanofibers forms at pH below 6 following a micelle-to-fiber phase transition. We then find that thixotropic hydrogels spontaneously form below pH 6 and of which the strength depends on the concentration and final pH, similarly to peptidic LMWG.[53,54] This behavior is different from what it was previously found for the deacetylated acidic C18:0 sophorolipid, for which fibrillation is a diffusion-limited process, with the gel strength depending on the pH change rate.[50] Finally, C16:0 sophorolipids hydrogels reach elastic moduli above 10 kPa for concentrations below 5 wt%, a range of values comparable to the best LMWG, such as FMOC derivatives.[55,56] Typically of thixotropic hydrogels, they also show fast recovery to about 80% of their mechanical strength within 20 s, and 100 % after 10 min, after being destructured with large strain amplitude beyond the linear domain during 5 min.

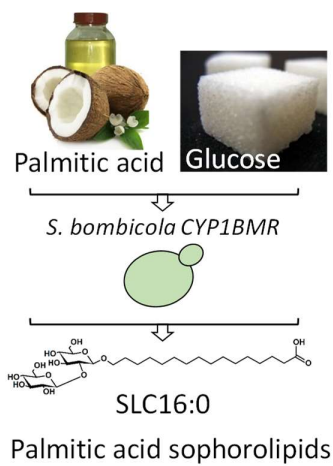

**Figure 1 – Acidic C16:0 sophorolipids are obtained by fermentation of *S. bombicola* CYP1BMR in the presence of palmitic acid and glucose. The compound is referred to SLC16:0 throughout this work.**

**Materials and Methods**

*Chemicals.* Palmitic acid sophorolipids SLC16:0 ($M_w$ = 596.7 g.mol$^{-1}$) were produced at a production rate of 33 mgL$^{-1}$h$^{-1}$ in a bioreactor system using a strain of the yeast *Starmerella bombicola*, modified by heterologous expression of the cytochrome P450 *cyp1* gene of *Ustilago maydis* and feeding with palmitic acid.[52] The molecule was hydrolyzed under alkaline conditions and only the non-acetylated acidic form of SLC16:0 was collected and used in this study. The detailed conditions of biosynthesis, purification as well as the full HPLC-ELSD, LC-MS and NMR characterization are provided in Ref. [52]. The typical $^1$H NMR spectrum of the compound used in this work is reported in Figure S 1 in the Supporting Information.

*General method to prepare SLC16:0 hydrogels.* SLC16:0 is generally insoluble in water at room temperature. However, the solubility of its deionized form at basic pH is improved, as classically observed for many glycolipid biosurfactants.[36,37,57] For this reason, SAFiN and hydrogels are prepared by a pH-jump process, from basic to acidic pH, following a procedure developed in previous studies.[35,37] The SLC16:0 solution (exact concentrations are given in the legends of figures) is adjusted to pH ~10 by adding 1-5 µL of NaOH 5 M. The solution becomes clear, indicating solubilization of the compound. pH is then reduced by adding 1-20 µL of HCl 0.5 M or 1 M (0.1 M can also be used for refinement). The exact amount of NaOH and HCl depends on the lipid concentration in solution. More precise data are given for the *in situ* SAXS experiments and can be found in Table S 1 the Supporting Information, which also provides the typical dilution factors and NaCl concentrations after the pH jump for a SLC6:0 system at C= 0.5 wt% acidified with either a 0.1 M or 0.5 M HCl solution. Fibrillation and gelation occur below pH ~6.

Additional information on the analytical techniques (*in situ* SAXS, analysis of SAXS data, rheology, cryo-TEM) is given in the Supporting Information.

**Results and discussion**

**Self-assembly of SLC16:0 in water**. The non-acetylated palmitic acid derivative of sophorolipids, SLC16:0, is a new molecule belonging to the broad family of biobased amphiphiles and its self-assembly properties in water are not known. Similarly to other sophorolipids, it contains a free-standing ionizable COOH group, of which the pKa is evaluated in this work at pKa ~7.0 by acido-basic titration (Figure S 2).

In this work, we study the aqueous phase behaviour of SLC16:0 at concentrations between 0.5 wt% and 5 wt% as well as its hydrogel-formation properties. The self-assembly is characterized by pH-resolved *in situ* SAXS (Figure 2, Figure 3, Figure S 3, Figure S 4), and by cryogenic Transmission Electron Microscopy (cryo-TEM, Figure 4, Figure S 5). The pH-resolved *in situ* SAXS experiment is designed according to previous studies: the sample solution (2 mL, 0.5 wt%) is pumped from the reaction beaker in a flow-through 1.5 mm quartz capillary by mean of a peristaltic pump. pH is controlled by adding microliter-amounts of a 0.5 M, or 0.1 M, HCl solution at a rate of 0.136 µL/s (Table S 1 for more information) by mean of a computer-controlled push-syringe. pH is monitored and acquired using a computer-controlled pH meter. pH and SAXS acquisitions are synchronized (1 acquisition every 5 sec) and triggered manually with an error of ± 1 s. [37]

The full SAXS profiles recorded from pH 9.6 to 2 are shown in Figure 2a for an added 0.5 M HCl solution and in Figure S 4a for a 0.1 M HCl solution. The advantage of the former is the negligible dilution factor (1.6 % against 17% for the 0.1 M HCl solution, Table S 1 for more information), although the rate of pH change is much faster than the latter, as pH 2 is reached within 13 min against 50 min when the 0.1 M HCl solution is employed (Figure S 4b). The rate of pH variation was shown to be important in the fibrillation process of the C18:0 sophorolipid congener, for which fast rates induce precipitation due to spherulite formation. [50] On the contrary, solutions of SLC16:0 fibers are always homogeneous and stable, independently of the molarity of the acid solution used and/or the rate of pH variation. This particular feature will be discussed later.

Two selected scattering curves, at pH 9.38 and 6.54, are given in Figure 2b and they display a similar signal, over an analogous q-range, to that observed for the fully ionized form at basic pH of other microbial glycolipids, sophorolipids, glucolipids and cellobioselipids [33,37,58]: an intense scattering below q= 0.01 Å$^{-1}$ and a broad oscillation of low intensity above q= 0.02 Å$^{-1}$. In analogy with the scattering signal of other microbial glycolipids, we qualitatively attribute the low-q scattering to aggregated objects and the oscillation to micellar aggregates. A more detailed description will follow in the next paragraphs.

The pH-driven experiment in Figure 2a shows that a phase transition occurs between pH 6.5 and pH 5.5, as indicated by the evolution in the scattering signal in the mid and low-q portion and by the appearance of a broad correlation peak at q= ~0.2 Å$^{-1}$ (Figure 2a, Figure 3a,b). Below pH 5.5, the signal has evolved into a strong, better defined, scattering contribution at low q and a well-defined diffraction peak at q= ~0.2 Å$^{-1}$.

The qualitative evolution of the SAXS profiles over the basic-to-acidic pH range has a similar behaviour to the previously-studied microbial glycolipids [33,37,58,59] and it identifies a complex behaviour where at least two phases coexist at the same time. For this reason, the analysis of the SAXS data is not straightforward and it was approached by two methods: a model-independent slope and peak analyses, presented in Figure 3, but also a model-dependent analysis of the form factor at q> ~0.03 Å$^{-1}$, presented in Figure S 3. The scattering data in the basic-to-neutral pH region hardly show a clear-cut Guinier plateau at low-q, thus preventing a clear-cut attribution of the origin of the low-q scattering. For this reason, we prefer to present the model-independent analysis in the main text and the model-dependent approach in the Supporting Information.

In the model-independent analysis, the value of the slope is related to q-dependence of the scattered intensity in the log(I)-log(q) representation of SAXS data. It can be characteristic of well-defined, although simple, morphologies (e.g., spheres, flat sheets, cylinders, sharp interfaces) [60] or more complex fractal systems. [61] This approach does not need any preliminary hypothesis on the morphology, an obvious advantage, although the final interpretation needs complementary data. The model-dependent analysis requires a hypothesis on the morphology and it eventually provides quantitative pieces of information on the structure (e.g., core-shell structure, size, thickness, density…). For the latter, we base our hypotheses on the data collected on similar systems [33,37,58,59] and on the fact that the low-q scattering could be associated to aggregated objects of ill-defined nature (aggregation of micelles or other morphologies, like platelets, as found elsewhere).[37,59]

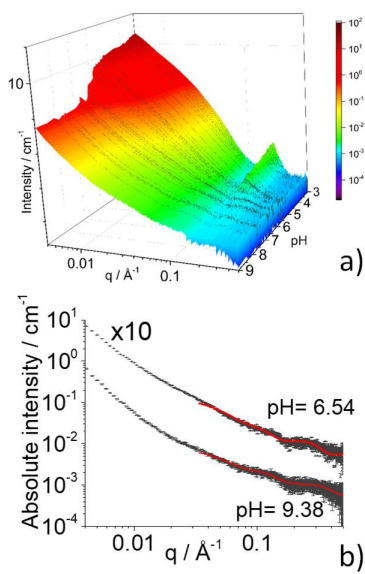

**Figure 2 - a) pH-dependent *in situ* SAXS profiles recorded for SLC16:0 solution (V= 2 mL) at T= 23°C and C= 0.5 wt% in water. pH is varied from basic to acidic by adding micromolar amounts of HCl solution (rate: 0.136 µL/s, total volume added, 32 µL, dilution factor= 1.6%, Table S 2 in the Supporting Information) at 0.5 M. b) Selected SAXS profiles. The profile at pH= 6.54 is shifted by a factor ten for clarity. Red lines correspond to the fit generated with a core-shell (prolate) ellipsoid of revolution model form factor. The model, presented in Ref. [62], and fitting results are presented in the Supporting Information. The results of the model-dependent fitting approach are shown and discussed in Figure S 3.**

The region below q< 0.03 Å$^{-1}$ is informative on the type of morphology at typical scales larger than about 200 Å. The q dependence of the intensity in the log-log scale, referred to as the slope, provides the fractal dimension, Df, a model-independent parameter which is either related to a specific morphology (e.g., -4 for spheres or sharp interfaces, -1 for cylinders, rods and elongated objects, -2 for lamellar and flat objects)[60] or describing the presence of bulk, or surface, fractal objects for intermediate values.[61] In the present work, the SAXS data in Figure 3a-c show two distinct scattering regimes at q< 0.1 Å$^{-1}$; for this reason, we analyze the values of the slope in two regions, below q= 0.03 Å$^{-1}$ (low-q regime) and between 0.03 < q / Å$^{-1}$ < 0.1 (mid-q regime, Figure 3a) for the entire pH range. The values of the slopes against pH for these regimes are reported in Figure 3c.

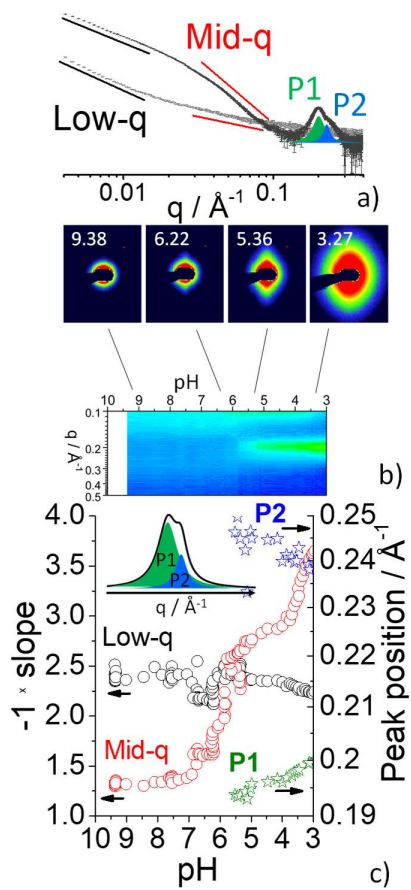

**Figure 3 – a)** Typical SAXS profiles extracted at pH> 7 (grey) and pH< 5 (black) from experiment in Figure 2a. Continuous black and red lines indicate the q-domains of linear fits in log-log scale while P1 and P2 schematically show the Lorentzian peaks used to fit the experimental diffraction peak above 0.1 Å⁻¹. A specific analysis of the structural features of P1 and P2 is given in the Supporting Information (Page S15). **b)** Contour plot profile of experiment in Figure 2a centered around the diffraction peak and selected, corresponding 2D SAXS images. **c)** pH-resolved evolution of low-q and mid-q slopes (left ordinate) and peak position (right ordinate).

The slope in the low-q region (black circles) is practically contained between -2 and -2.5 throughout the entire pH range. The non-integer values indicate the presence of fractal objects but its proximity to -2 strongly suggests that the morphology of single objects is planar. The slope in the mid-q region (red circles) evolves, on the contrary, between close to -1 (pH 9 to 7) and -3.5 (pH 3). At pH> 7, one can reasonably make two hypotheses: coexistence between the micellar phase and a second phase, presumably composed of platelets or bilayer fragments, as it was found for other acidic sophorolipids and glucolipids,[37,59] or aggregation of the micelles at scale larger than about 600 Å.

Within this framework, one can push further the analysis of the broad oscillation above about 0.1 Å$^{-1}$ and use a model-dependent approach to estimate the size and structure of the micellar objects. This is shown in Figure S 3 and a detailed discussion can be found in the Supporting Information. As a short summary, micelles could be described as ellipsoidal objects with an equatorial radius of 13 ± 1.3 Å and a polar dimension of 49 Å ± 4.9 Å above pH 7. Below pH 6, that is upon increasing content of the COOH form of SLC16:0, the equatorial and polar radii respectively increase at 26 ± 2.6 Å and 72 ± 7.2 Å (Figure S 3b). In the present model, we assume a core electron density, constituted by the aliphatic part of SLC16:0, and shell density, constituted by sophorose, the COOH group and water. Considering the coexistence of COOH and COO$^{-}$, of which the ratio vary with pH, we suppose that the electron density distribution of the shell is not homogeneous. To rationalize such heterogeneity, we employ a model form factor where the shell thickness is not homogeneous.[63] This model is certainly an approximation, but it is the only one which helps evaluating at best the fluctuation in electron density around the hydrophilic shell associated with the coexistence of the carboxylic and carboxylate forms of SLC16:0.

When pH is between 5 and 6, the mid-q slope reaches the value of -2, being the same range as the low-q slope. This behavior strongly suggests a continuous morphological change from ellipsoidal to flat objects. When the pH falls below 5.5, the cross section varies only by not more than 10 Å, as indicated by the contained evolution of the minimum of the form factor from 0.2 Å$^{-1}$ to 0.15 Å$^{-1}$, the cross-section is most likely flat, as indicated by the slope around -2 at low-q but the length increases monotonously much above 150 nm, beyond the detection window of the present SAXS configuration, as indicated by the slope settled around -3.5 (interface) in the mid-q and by the lack of a plateau at low-q. The evolution of the SAXS signal below 5.5 is then typical of either anisotropic fibrillation or formation of flat lamellae.

This hypothesis is strongly supported by the anisotropic scattering signal in the 2D SAXS images recorded below pH 6.25 (Figure 3b). These are typical of anisotropic structures aligned in the flow direction within the capillary and orthogonal to the direction of the incident beam. The flat anisotropic structures are also characterized by a broad diffraction peak at q= ~0.20 Å$^{-1}$, indicating the formation of a crystalline order. The peak can be actually deconvoluted into two Lorentzian contributions (Figure 3a,c), P1, evolving from 0.19 Å$^{-1}$ at pH 5.5 to 0.2 Å$^{-1}$ at pH 3, and P2, evolving from 0.25 Å$^{-1}$ to 0.24 Å$^{-1}$ in the same pH range. A more detailed analysis of the evolution of the structural features (position, full width at half maximum and intensity) of P1 and P2 is given in the Supporting Information (Page S15). All in all, as discussed later, the peak centered at about 0.19 Å$^{-1}$ seems to be the most important in the final material, irrespective of the synthesis conditions.

A broad diffraction peak at a comparable q-value is commonly observed for other biosurfactants' systems. It has been attributed to the typical inter-lipid distance laying in the plane of twisted nanofibers[35] but it could also correspond to the intermembrane repeating distance in liquid crystalline lamellar phases stabilized by repulsive electrostatic interactions.[40,64] We anticipate that the second hypothesis is ruled out both by cryo-TEM arguments and by the insensitivity of the peak position to increasing ionic strength. In conclusion, despite such modest variations, we do not observe any other specific differences (form factor, position of minima, low-q or mid-q slopes) between a faster or slower acidification rate. We then conclude that this parameter has a negligible influence on the structure of SLC16:0 structures within the length scale explored by SAXS.

The nature of the self-assembled structures at acidic pH has been studied by complementary cryo-TEM experiments, presented in Figure 4 and Figure S 5, and recorded at pH 3 and at the concentration of 0.25 wt%, so to avoid overload of the TEM support grid. At low magnification (Figure 4a, Figure S 5a), the sample is massively constituted by "infinitely" long fibers, organized in bundles and often aligned in a given direction, thus explaining the anisotropic signal found in SAXS experiments (Figure 3b). A closer look (Figure 4b) shows the presence of flat structures, with an average diameter of 8.6 ± 0.9 nm, that is about 10% polydispersity, measured over about 50 different fibers. Interestingly, tilting of the TEM sample holder (+33°, +35° and +40° tested) shows a homogeneous electron density across the fibers' section, as indicated by the arrows 1 through 3 at 0° (Figure 4c) and at +33° (Figure 4d). Estimation of the cross-section on tilted sample confirms an average diameter of 8 nm. This curious result could suggest that the fibers are actually cylinders, or nanotubes. However, these morphologies would provide a -1, or close to -1, slope in the low-q portion of SAXS experiments at pH< 6 and this is in contradiction with all our SAXS data, showing an approximate -2 slope, instead (Figure 3c), and typical for flat structures.[35,60,65,66] The structures in Figure 4 a-c could then be compatible with a nanobelt, also displaying a -2 slope in SAXS,[19] but a difference in the diameter of the cross-section would be expected upon tilting the sample holder.[19] This is not the case here (Figure 4c,d). Flat ribbons then constitute the only plausible structure. This hypothesis would agree with the self-assembled morphology of the analogue stearic C18:0 sophorolipid,[35] although the typical twisted ribbon structure is not frankly visible in our TEM images. The possible combination of long pitch values, thin cross-section (< 10 nm), poor contrast and limited resolution of our TEM camera could probably explain such discrepancy.

Interestingly, cryo-TEM also reveals the presence of large flat crystals, of which the extremities disassembles into fibers. This is illustrated in the close-up of Figure 4e and in Figure S 5b-e. The Fourier transform (panel 2-FT) of region 2 and plot profile 1 of Figure 4e show a highly crystalline region with repeating distance of 2.63 nm. Plot profiles 2 through 4 in Figure S 5c-e show additional crystalline regions of a typical interplanar distance of 3.33 nm, 2.75 nm and 2.67 nm. These values are in agreement with positions of P1 (0.19-0.20 $\text{Å}^{-1}$ ≡ 3.31-3.14 nm) and P2 (0.23-0.25 $\text{Å}^{-1}$ ≡ 2.73-2.51 nm). Both SAXS and cryo-TEM then suggest the simultaneous existence of several structural polymorphs in terms of lipid organization within the fibers, as explained below.

The d-spacing values found here are all typical for lipid nanoribbons, nanobelts and nanotubes,[16,19,67] although the size of the molecule and its packing have a major importance.[67–70] Masuda and coworkers have described in detail the correlation between the position of the strongest reflection in lipid nanotubes formed by unsymmetrical bolaamphiphiles and the type of polymorph and polytype in relationship to the estimated length of the molecule in an all-*trans* configuration.[67] In the case of multiple peaks in a narrow q-range around q= 0.2 $\text{Å}^{-1}$, the coexistence of more than one polymorph and polytype is then not to be excluded. The calculated length of the SLC16:0 is approximately 31 Å, where about 21 Å are related to palmitic acid, the length of which is estimated with the Tanford formula (1.54+1.265*$n$, $n$= number of $CH_2$ groups),[71] and 10 Å is a typical size for a disaccharide.[72] If this calculation shows that the d-spacing is in the order of the molecular length, the coexistence of two main distances centered around 32 Å and 26 Å suggests the presence of both untilted and tilted polymorphs.[16,67]

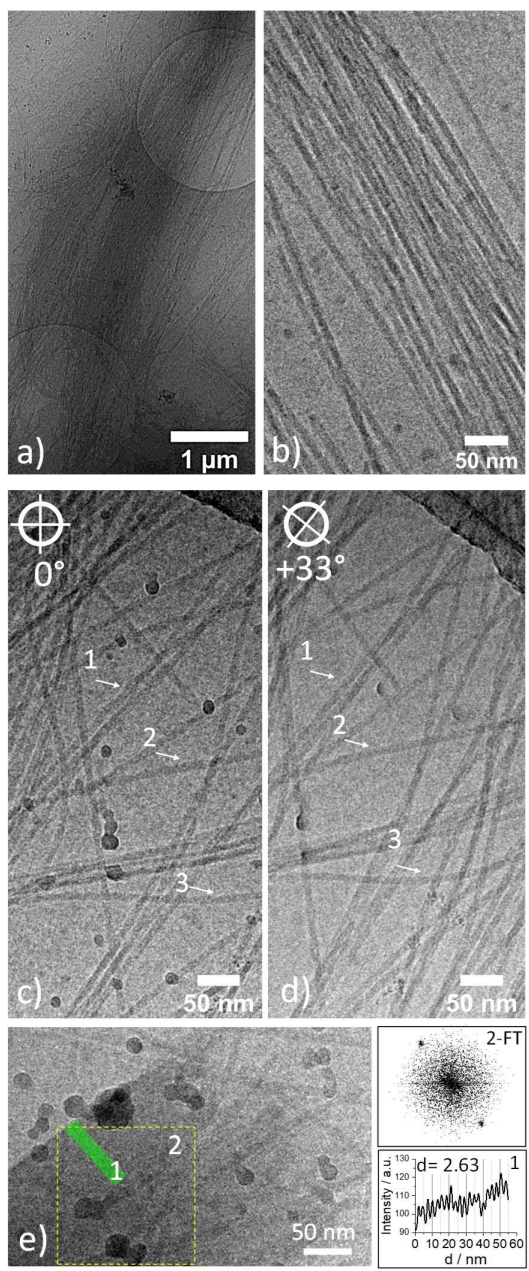

**Figure 4 – a-e) Cryo-TEM images recorded on a SLC16:0 sample at C= 0.25 wt% and pH 3. Sample is diluted 10 times from a hydrogel prepared at C= 2.5 wt% using the pH jump approach (pH 10 → pH 3). Images have been analyzed with Fiji software.[73]**

The mechanism of formation of SLC16:0 fibers undergoes a continuous micelle-to-fiber transition, characterized by a morphological evolution of the micelles. When the pH decreases, the hydrophilic shell becomes more homogeneous in the equatorial direction, most likely due to the increasing amount of carboxylic acids. Micelles then fuse together both in the equatorial and polar directions, probably through H-bonding interactions, with a morphological evolution from spheroids to flat objects, which crystallize below the pKa of SL16:0. Crystallization seems to occurs on large scale lengths, of a few hundred nanometers, although such structure eventually peel into highly homogeneous fibers, of possible ribbon morphology, of about 8 nm in cross-section with 10% polydispersity. A similar peeling mechanism was reported before for peptide and lipid nanobelts as a function of concentration.[19] If peeling is possible and actually observed by cryo-TEM, it is at the moment unclear whether or not individual fibers can also grow directly from the micelles. This hypothesis cannot be excluded and it is probably concomitant with a peeling mechanism.

Interestingly, the self-assembly of SLC16:0 strongly differs from that of SLC18:0, which undergoes an abrupt micelle-to-ribbon transition with no apparent morphological evolution based on *in situ* SAXS.[37] Fibrillation was then explained by a classical nucleation and growth mechanism where the micellar phase is seen as a reservoir of matter only, from which molecules diffuse to the nuclei. On the contrary, SLC16:0 micelles seem to behave as growth sites for the nanofibers, as reported for amyloid fibrillation.[74] A possible explanation of such discrepancy will be given in the discussion section.

**Hydrogel from SLC16:0 SAFiN**. From a minimal concentration of 0.5 wt%, the SLC16:0 solutions spontaneously form a gel when pH is lowered below ~6. To confirm that concentration does not modify the fibrillar structure, we perform complementary SAXS on the SLC16:0 hydrogels. The typical SAXS profiles recorded at higher concentrations (up to 2.5 wt%), given in Figure 5a, show similar features as the ones recorded at a lower concentration (Figure 3a): a -2 slope at q < 0.03 Å$^{-1}$ is associated to a main diffraction peak at q= 0.194 Å$^{-1}$. For these reasons, the concentrations explored do not have any impact on the nature of the self-assembled fibrillar structures. Interestingly, the main diffraction peak for concentrated gels settles around 0.19 Å$^{-1}$. First of all, this feature tells that this specific peak position, corresponding to P1 in Figure 3, is eventually favored over P2. Secondly, it also indicates that its corresponding repeating distance of 32.4 Å, hence the untilted polymorph, compared to the expected length of SLC16:0, is the preferred arrangement of SLC16:0 molecules within the fiber. Interestingly, the peak at double its value, q= 0.388 Å$^{-1}$, suggests a lamellar order of SLC16:0 within the ribbon plane.

Salt (NaCl) is generally generated during the pH jump approach. In the case of SAFiN obtained from sophorolids, NaCl was shown to influence the homogeneity of the fibers' cross-section.[46] In the case of lamellar liquid crystalline systems, included hydrogels, where electrostatic repulsions prevent the lamellar structures from collapsing, salt is responsible for screening the long-range electrostatic forces. This is experimentally followed by a shift of the lamellar peak towards higher q-values.[40,75] In the present work, the amount of salt generated during the pH jump process for a 0.5 wt% SLC16:0 solution is evaluated to 7.9 mM and 17.2 mM when the 0.5 M and 0.1 M HCl solutions are respectively employed (Table S 1 in the Supporting Information). These values are low and, considering the similarities among the structural parameters evaluated with the pH-dependent *in situ* SAXS experiments, one can safely state that NaCl does not have a major impact. However, to dissipate any doubt, we have performed additional SAXS experiments on hydrogel samples to which NaCl is deliberately added. Data presented in Figure S 6 show a substantial similarity among all scattering profiles in the entire q-range explored, thus confirming that NaCl does not have any unexpected structural effects on the SLC16:0 fibers up to 250 mM NaCl. It is also interesting to note that the diffraction peak at q= 0.19 Å$^{-1}$ is insensitive to ionic strength, which excludes inter-fibers repulsive electrostatic interactions and confirms the attribution of this peak to an intra-fiber long-range order.

The viscoelastic properties presented below are studied at T= 20°C on a series of samples freshly prepared in water using the pH jump approach. Each sample has been systematically vortexed to remove its shear history, before loading it in a plate-plate geometry. The frequency-sweep experiments recorded in the linear viscoelastic regime (LVR) are performed after 30 min from loading. Frequency-sweep experiments are followed by strain sweep (Figure S 7) and step-strain experiments (Figure 6 and Figure S 8). The latter is employed to evaluate the recovery potential of SLC16:0 hydrogels after applying shear stress with an amplitude out of the LVR. Studying the rheological behavior of the gels after 30 min from loading has the advantage of limiting the dehydration, although its main drawback consists of the fact that the mechanical properties could still evolve in time. If our choice may impact the absolute values of G′, it does not influence the general trends and relative comparison among the samples at a given time.

Figure 5b shows the typical frequency sweep experiments recorded in the LVR (γ= 0.1%, Figure S 7) for a set of samples at 2.5 wt% and pH between 3 and 6, while Figure 5c reports the concentration-dependent G′ values recorded at pH 5. G′ and G′′ curves are parallel with G′> G′′ in the entire frequency range, thus demonstrating the presence of a gel for all pH. The sample at concentration as low as 0.5 wt% has an elastic modulus, G′, of about 45 Pa, thus showing that SLC16:0 forms hydrogels at concentrations even below 1 wt%. The elastic modulus scales linearly with concentration (Figure 5c), with a slope of 3.14 ± 0.03. This value lies between typical exponents found in weakly aggregating polymer colloids (3.7 and 4.5)[76] and entangled polymer, biopolymer and fibrillary hydrogels (~2.3).[44,50,53,77–81] Such discrepancy was observed before for similar molecules and it is correlated to the gel equilibration time.[44] The storage modulus measured at concentrations between 2.5 wt% and 5 wt% and pH 5 varies between 4 kPa and 40 kPa, a range which is consistent with other SAFiN hydrogels,[54,81,82] including sophorolipids.[50]

pH has a crucial effect on the elastic properties (Figure 5b). In the vicinity of the micellar-to-fiber phase transition at pH 6, the elastic modulus at 2.5 wt% is in the order of 1 kPa, while it increases at 7 kPa at pH 3 for the same concentration. As far as the ionic strength is concerned, tested between 0 mM and 250 mM, we did not observe any impact on the SLC16:0 mechanical properties. The strong impact of pH is most likely related to the massive fibrillation region below the pKa. Closer to the pKa, the content of carboxylate molecules and the volume fraction of micelles is still too high to obtain the strongest hydrogels. However, upon pH decrease, the content of carboxylic molecules is maximized, as well as the volume fraction of fibers. Such a strong connection between final pH and mechanical properties in SAFiN constituted of LMWG is not uncommon and it was described before for FMOC peptides.[53,54] However, this mechanism strongly differs from the one found for C18:0 sophorolipids, for which the final pH has no impact on the mechanical properties, which are instead controlled by the pH change rate.[50] Such difference will be commented further.

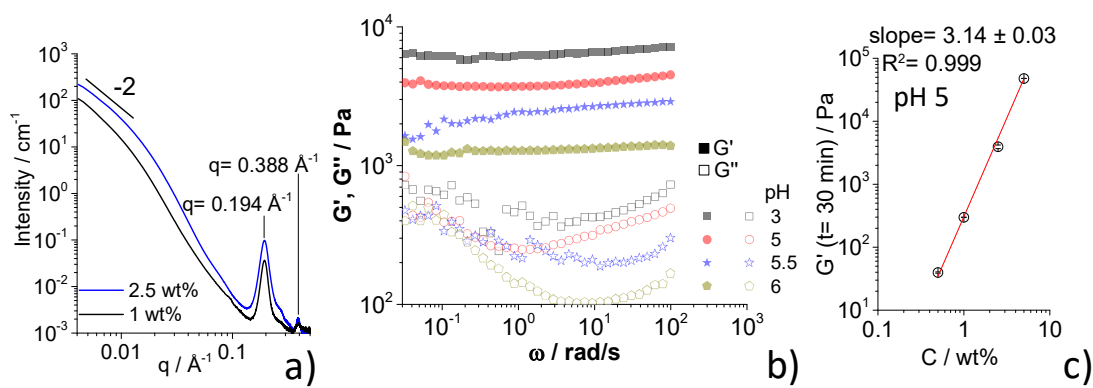

**Figure 5** – SAXS experiments recorded on SLC16:0 hydrogels at C= 1 wt% and 2.5 wt%, both at pH 5. b) Frequency sweep profiles of storage (G') and loss (G'') moduli recorded in the linear viscoelastic regime (γ= 0.1 %, T= 20°C) for a series of SLC16:0 samples prepared at C= 2.5 wt% and various pH. c) Concentration dependency of G' measured in the linear viscoelastic regime after 30 min from sample loading (ω= 6.28 rad/s, γ= 0.1 %, T= 20°C).

After application of an oscillatory strain with amplitude out of the LVR, SLC16:0 hydrogels show a time-dependent recovery of their elastic modulus, typical of thixotropic gels. The recovery properties of the gels are tested against oscillatory strain amplitude applied outside the LVR ($\gamma$= 100%). During 120 min, SLC16:0 hydrogels undergo a series of harsh step-strain cycles, of which six consecutive ones were designed with $\gamma$= 100% during 5 min followed by 10 min recovery in the LVR. Figure 6a and Figure S 8 show a series of step-strain experiments performed on SLC16:0 hydrogels at pH 3, 5 and 6 and C= 2.5 wt%. The restructuring process, which lasts over several minutes, is quantified by comparing the % of $G'$ recovery after 20 s and 5 min at the $N^{th}$ plateau with respect to the N-1 plateau and with respect to the zero-plateau (N= 0). This is illustrated in Figure 6a,b for the system at pH 5 while the % of recovery is given in Figure 7 for all pH values.

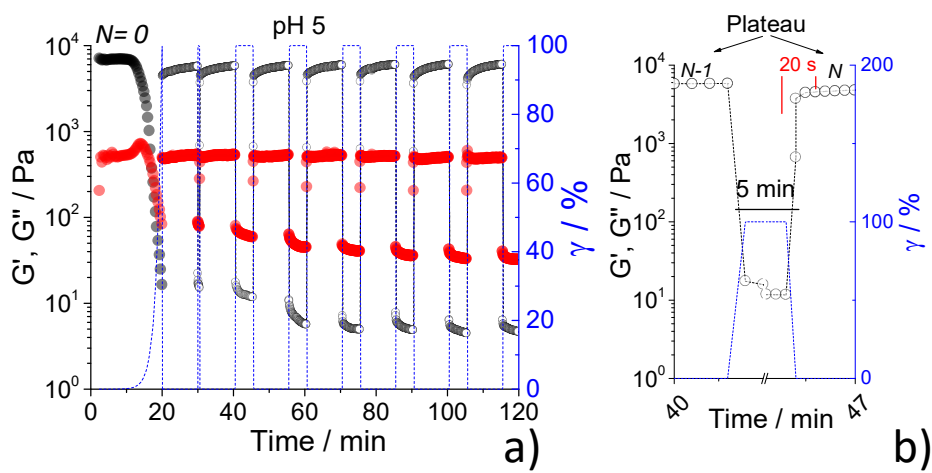

**Figure 6 – Study of the recovery properties of SLC16:0 hydrogels (C= 2.5 wt%, pH 5, $\omega$= 6.28 rad/s) after a series of imposed oscillatory shear strain with amplitude out of the linear viscoelastic domain. a) Evolution of $G'$ and $G''$ in a typical step-strain experiment performed with an initial shear strain (logarithmic evolution, $4.10^{-3} < \gamma < 100$ %) followed by a recovery of 10 min at $\gamma$= 0.1%. Six cycles of step strain experiments then follow: the first cycle consists of applying a strain of $\gamma$= 100% during 30 s followed by a recovery of 10 min; cycles form 2 to 6 consist of applying a strain of $\gamma$= 100% during 5 min followed by a recovery of 10 min. b) Specific highlight between the second and third cycle showing the region concerned by the recovery time of 20 s.**

Before beginning the first cycle, the value of $G'$ at plateau at pH 5 is at about 6900 Pa and $G' \gg G''$ (more than one decade). A strain amplitude at $\gamma$= 100% is progressively applied, during which $G'' \gg G'$, confirming the complete destructuring of the gel. When the strain is set again in the LVR, $G' \gg G''$ within 5 s, with a recovery of 65 % after 20 s and 83 % after 10 min. Figure 6a also shows that longer (5 min) and repeated (6 times) destructuring actions do not have a significant negative impact on the hydrogel's mechanical properties. This is highlighted in Figure 6b for the second step-strain action and on the evolution of the % of recovery averaged over 6 cycles, shown in Figure 7. Whichever the pH in the sample, the recovery with respect to the initial, N= 0, plateau (red columns) settles between 60% and 70% after 20 s and between 80% and 90% after 10 min. The value found above 100% at pH 6 after 10 min is probably due to the continuous evolution of this gel. Figure 7 also shows that the recovery after 20 s with respect to the N-1 plateau (grey columns) reaches an average of 78% at pH 3 and pH 5 and 62% at pH 6. After 10 min, the recovery is 100% for all pH values.

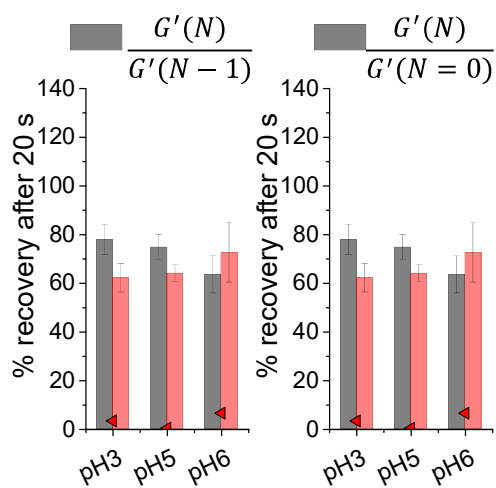

**Figure 7 - Percentage (%) of recovery after 20 s and 5 min for the experiment in Figure 6a). Grey bars show the recovery with respect to the N-1 plateau, while red bars with respect to the N= 0 plateau.**

**Discussion**.

The combination of pH-resolved *in situ* SAXS and *ex situ* cryo-TEM shows that a micelle-to-fiber transition occurs in the vicinity of the pKa of SLC16:0. Fitting of the SAXS profiles using a core-shell ellipsoid of revolution form factor suggest that the equatorial shell region of the micelles becomes more and more occupied by the COOH groups when pH approaches 7, below which a morphological change occurs from spheroidal to flat objects. Redistribution of the COOH groups in the equatorial region could enhance lateral H-bonding interactions between adjacent micelles and drive the formation of flat fibers, which eventually in the polar, longitudinal, direction. Growth in the equatorial, lateral, direction is not excluded, although, in this case, individual fibers eventually peel off, as shown by cryo-TEM. Although the crystalline packing of SLC16:0 within the fibers can adopt several polymorphs, it seems that the untilted arrangement, with a diffraction peak at about 0.19 Å⁻¹, constitute the equilibrium distance at the end of the fibrillation process. However, neither the exact polymorph (symmetrical or unsymmetrical) nor the exact polytype (head-to-head, head-to-tail) in terms of the respective bolaamphiphile arrangement within the fibers, as intended by Masuda *et al*.,[67] is known with exactitude at the moment. Enhanced fibrillation below pH 6 is responsible for the spontaneous formation of hydrogels with very good mechanical properties, displaying elastic moduli above 10 kPa. The most probable micelle-to-fiber-to hydrogel mechanism is summarized in Figure 8.

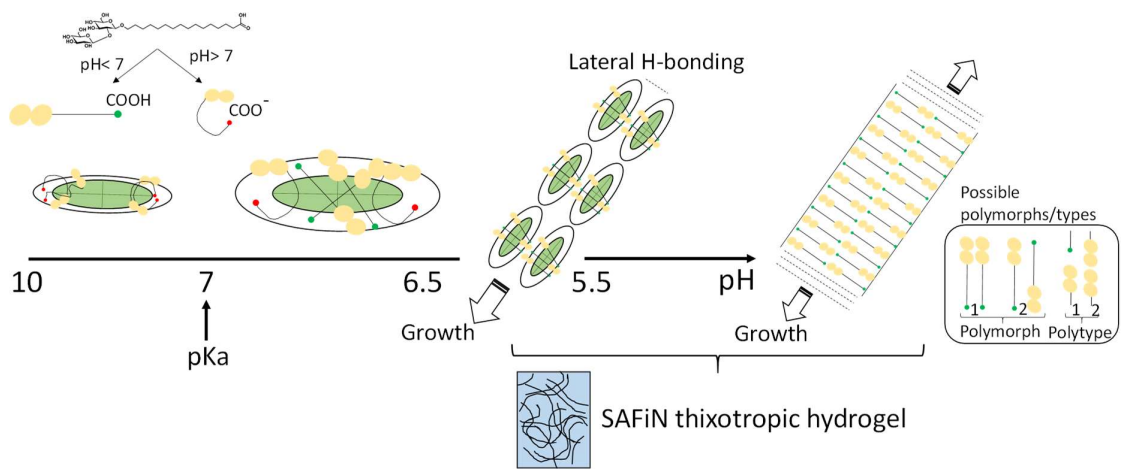

**Figure 8 – Proposed pH-dependent mechanism of fibrillation and SAFiN thixotropic hydrogel formation for palmitic acid C16:0 sophorolipids. Possibilities of polymorphs 1 and 2 and polytypes 1 and 2 subsist.**

The fiber phase found in the SLC16:0 system at pH< 6 seems to be at equilibrium at room temperature. The fibers' cross-section diameter is highly homogeneous (10% polydispersity), compared to the corresponding SLC18:0 nanoribbons. The latter, which differ only in two methylene groups, fibrillate below pH 7.4, but the distribution of the fibers' diameters was shown to vary between 8-10 nm and up to 20-30 nm.[35] Salt was shown to play an important role in the size dispersion, the lower the salt content, the narrower the size distribution.[46] In the meanwhile, salt was also observed to favour spherulite formation over homogeneous fibers and precipitation over gel formation for the same C18:0 sophorolipid molecule. Overall, the hydrogels' mechanical properties of C18:0 sophorolipids were strongly affected by the rate of pH variation.[50] The strong sensitivity of the mechanical properties of SLC18:0 fibrillar hydrogels to pH change rate and ionic strength are most likely correlated to its abrupt micelle-to-fiber transition. pH-resolved *in situ* SAXS experiments have shown that micelles do not change their morphology into fibers upon increasing the COOH content at lower pH, but it was supposed that fibers nucleate in solution when an excess of carboxylic acid C18:0 sophorolipids are formed, whereas micelles only constitute a reservoir of matter. The fibrillation (nucleation and growth) process is then controlled by diffusion of the acidic C18:0 sophorolipids from the micellar environment to the fibers.[37,50] Diffusion-controlled fibrillation is well-known for temperature-stimulated LMWG, for which it was shown that the rate of temperature variation controls the molecular solubility equilibrium. Fast temperature (or pH, in the case of SLC18:0) variation favors supersaturation of the insoluble component, thus driving spherulite formation and, eventually, poor mechanical properties of the hydrogel. On the contrary, slow temperature (or pH) variation favors equilibrium between soluble and insoluble species, thus minimizing the mismatch nucleation energy and, consequently, spherulite content. In the latter case, more homogeneous fibers and stronger gels are eventually formed.[51,83]

The fibrillation mechanism seems to be completely different in the case of SLC16:0, which shows a continuous morphological evolution between micelles and fiber upon acidification of the medium (Figure 2, Figure 3). In this case, fibrillation probably does not follow the nucleation and growth process controlled by the diffusion of the molecular SLC16:0 species in solution. In this case, hydrogel stability and strength do not depend on the rate at which fibrillation occurs, but rather on the overall volume fraction of the fibers in solution. If one assumes that the fibers are most likely composed of the carboxylic acid derivative of SLC16:0 (highly possible but not demonstrated at present), their volume fraction only depends on the final pH. This is experimentally verified through pH-dependent rheology experiments (Figure 5b). These properties are consistent with what it is classically reported for FMOC-based peptides.[53,54]

Why do SLC16:0 and SLC18:0 sophorolipids, which only differ in two methylene groups, undergo a similar micelle-to-fiber transition with decreasing pH, but result in hydrogels with different sensitivity to pH? The answer could be found in the different melting-crystallization profiles between the C18:0 and the C16:0 derivatives of sophorolipids (Figure S 9a). The former shows a glass transition temperature, $T_g$, at about 56°C and two well-defined first-order transitions at 77°C upon heating (fusion) and cooling (crystallization). This value is not much different than the melting temperature of stearic acid (69°C). On the other side, SLC16:0 shows no first-order transition in the 10°C to 110°C temperature range but only an ill-defined second-order transition between 85°C and 100°C (Figure S 9b). Considering the fact that the melting temperature of palmitic acid is 63°C, it is highly unlikely to expect first order transitions (fusion/crystallization) of SLC16:0 above 110°C. This suggests that acidic SLC16:0 molecules are in a more fluid state at room temperature, a fact which could explain the continuous micelle-to-fiber transition rather than abrupt crystallization.

SLC16:0 form fibrillar hydrogels, as observed for other synthetic[25,84] and microbial[44,48,50] glycolipids. SLC16:0 hydrogels have very interesting rheological properties, which can probably be explained by the good homogeneity of the fibrillar network. They can reach elastic moduli above 10 kPa below 3% and have thixotropic properties with fast recovery. For the latter, they show an average of 60% and 80% recovery, respectively after 20 s and 5 min, of their initial storage modulus, while they show a 80% and 100% recovery, respectively after 20 s and 5 min, of their storage modulus immediately preceding the applied stress. Thixotropic hydrogels with fast and complete recovery of the viscoelastic properties are interesting systems, especially for biomedical applications.[27] Specific peptide amphiphile hydrogels in the presence of HCl were submitted to a similar step-strain cycle as done in this work (5 min at 100% strain) and have shown recovery of 90% of the initial elastic modulus within 10 min,[81] while the dipeptide 2NapFF was shown to recover 100% of its initial modulus after the first shear deformation and an average of 58% after five cycles.[53] In this regard, peptide amphiphiles are considered as one of the most performing LMWG gelators in the literature, both in terms of absolute value of the elastic moduli and recovery after shear. The performances of SLC16:0 hydrogels can be certainly compared to the ones of this class of molecules.[85–87]

If homogeneous fibrillation, absence of spherulites and low polydispersity of the fibers' cross section can explain the hydrogel's properties, some open questions still remain. In particular, it could be interesting to know whether or not the fiber morphology (twisted, helical, tubular, belt-like…) has any impact on the macroscopic rheological properties. The extensive amount of work published in the field of physical self-assembled gels broadly addresses the structure as "fibrillar", often disregarding their morphology or the relationship between the cross-section size distribution and elastic moduli. This is understandable because low molecular weight gelators rarely self-assemble into a homogeneous well-defined structure with monodisperse diameter and length, but they are rather characterized by a complex network of poorly-defined fibers with tip and/or side branching, greatly affecting the mechanical properties.[51,82,88] In addition, the mechanisms of gel destructuring and recovery upon application of a mechanical stress are far from being trivial and may actually depend on the fiber aggregation state but also on fibers' breaking events. For instance, common sense suggests that application of shear stress to a SAFiN gel results in fiber alignment, similarly to what we observe in this work during the pH-resolved *in situ* study (Figure 3b). However, Pochan *et al.* have actually shown by mean of rheo-SANS experiments the lack of fibrillar alignment but rather the formation of fractured fibrillar domains allowing the gel to flow, although it was not clear whether or not the initial structure of the gel was constituted by a spherulitic, or fibrous, network. This is an important detail, which has a tremendous importance in the macroscopic behavior of the gel under shear flow and recovery.[85]

Unfortunately, fibrillar gels with very interesting recovery rates[85–87] are only partially characterized from a morphological perspective. Techniques like TEM or SEM are commonly used to characterize the fibrillar morphology. Still, these have several drawbacks: both of them require the drying of the sample and its study under vacuum, conditions which may strongly affect the real structure in solution; SEM only provides a surface survey but not an internal insight of the fibers. In addition, standard SEM rarely have the required resolution to probe local structures in the sub-10 nm range. Finally, in the absence of complementary techniques like cryo-TEM, even small angle neutron or X-ray scattering does not provide a complete structural resolution, because fibrillar systems generally show common features characterized by a -2 slope in the low-q regime and a structure peak above 0.1 $Å^{-1}$.

**Conclusion**

This work shows that palmitic acid C16:0 sophorolipids spontaneously self-assemble into micelles at pH above 7 and into fibers at pH below 6. pH-resolved *in situ* SAXS experiments show a structural continuity from the micelles to the fiber morphology, differently than what it is found for the congener C18:0 sophorolipids, which otherwise displays an abrupt transition between similar structures. The transition occurs below the pKa (7) and it seems to be driven by a redistribution of the carboxylic acids groups of SLC16:0 in the equatorial hydrophilic shell region of the micelles. The lack of first-order, melting or crystallization, transition between 10°C and 110°C on the solid powder supports the softer nature of SLC16:0 structures at room temperature and the possible absence of a nucleation and growth fibrillation mechanism when lowering pH, contrarily to what was found with the C18:0 sophorolipid congener. Lack of spherulitic structures in cryo-TEM also support this hypothesis. The advantage of the structural continuity between micelles and fibers is put in evidence by the homogeneity in terms of cross sections of the fibers, which have a diameter of average diameter of 8.6 ± 0.9 nm and infinite length.

These structural features directly impact the elastic properties of the material at concentrations above 0.5 wt%, when fibrillar hydrogels form spontaneously below pH 6. The elastic modulus reaches values as high as G'= 40 kPa at 5 wt% at pH 5, with final pH having a substantial impact: at 2.5 wt%, the elastic modulus increases by a factor 5 from pH 6 to pH

3. This behaviour is in agreement with FMOC peptides-based self-assembled hydrogels and in disagreement with C18:0 sophorolipid fibrillar hydrogels. The origin of such discrepancy is attributed to the different fibrillation mechanism. Finally, SLC16:0 hydrogels show thixotropic properties with fast recovery after being submitted to large oscillatory strain amplitudes for as long as 5 min. It is shown that, after removing the mechanical constraint, 80% of their elastic modulus is recovered after 20 s and 100% after 10 min, when comparing to the G′ values immediately preceding the constraint.

In summary, considering the ease of their synthesis procedure, strength, stability and fast recovery, SLC16:0 hydrogels have high potential for the development of sophorolipid-based soft fluids and materials.

**Acknowledgements:** This work benefited from the use of the SasView application, originally developed under NSF award DMR-0520547. SasView contains code developed with funding from the European Union's Horizon 2020 research and innovation programme under the SINE2020 project, grant agreement No 654000

**Funding**: European Community's Seventh Framework Programme (FP7/2007–2013) under Grant Agreement No. Biosurfing/289219; VLAIO - Agentschap Innoveren & Ondernemen, APPLISURF project. Diamond synchrotron radiation facility (Oxford, UK) has funded the access under the proposal N° SM23247-1; Soleil synchrotron radiation facility (Saint-Aubin, France) has funded the access under the proposal N° 20190961 and N° 20201747.

# Supplementary Information for

## Palmitic Acid Sophorolipid Biosurfactant: From Self-Assembled Fibrillar Network (SAFiN) To Hydrogels with Fast Recovery

Niki Baccile,[1,*] Ghazi Ben Messaoud,[1,†] Patrick Le Griel,[1] Nathan Cowieson,[2] Javier Perez,[3] Robin Geys,[4] Marilyn de Graeve,[4] Sophie L. K. W. Roelants,[4,5] Wim Soetaert[4,5]

[1] Sorbonne Université, Centre National de la Recherche Scientifique, Laboratoire de Chimie de la Matière Condensée de Paris, LCMCP, F-75005 Paris, France

[2] Diamond Light Source, Harwell Science and Innovation Campus, Didcot, Oxfordshire, OX11 0DE, UK

[3] Synchrotron Soleil, L'Orme des Merisiers, Saint-Aubin, BP48, 91192 Gif-sur-Yvette Cedex, France

[4] Ghent University, Centre for Industrial Biotechnology and Biocatalysis (InBio.be), Coupure Links 653, Ghent, Oost-Vlaanderen, BE 9000

[5] Bio Base Europe Pilot Plant, Rodenhuizekaai 1, Ghent, Oost-Vlaanderen, BE 9000

*Correspondence to: Dr. Niki Baccile, niki.baccile@sorbonne-universite.fr, Phone: 00 33 1 44 27 56 77

† Current address: DWI- Leibniz Institute for Interactive Materials, Forckenbeckstrasse 50, 52056, Aachen, Germany

**Supplementary Information**

**Materials and Methods**

*Small Angle X-ray Scattering (SAXS).* SAXS experiments were performed on two beamlines.

pH-resolved *in-situ* SAXS experiments were performed at T= 23 ± 1°C at the SWING beamline of SOLEIL synchrotron facility (Saint-Aubin, France). We used a flow-through quarts 1.5 mm capillary connected to the sample-containing solution at pH 10.4 through a peristaltic pump. The pH was controlled *in situ* via a KCl microelectrode (Mettler Toledo) connected to a computer-connected pH-meter (Hanna Instruments) located in the experimental hutch. pH is monitored in real time from the control room through a computer interface using the software provided by the manufacturer (Hanna instruments). pH changes have been obtained using a 0.5 M, or 0.1 M, HCl solution introduced at a rate of 0.136 μL/s with a computer-controlled press-syringe available at the beamline. Acquisition of pH and SAXS spectra is synchronized every 5 s and it is simultaneously triggered by hand with an estimated error of 1 s. The effective acquisition time of the SAXS spectra is 1 s, followed by 4 s delay time. The energy of the beam is set at 12 KeV and a sample-to-detector distance of 2.005 m. The signal of the Eiger 4M (Dectris) detector, used to record the data, is integrated azimuthally at the beamline using the software Foxtrot and in order to obtain the *I(q) vs. q* spectrum ($q = \frac{4\pi \sin\theta}{\lambda}$, where $2\theta$ is the scattering angle) after masking systematically wrong pixels and the beam stop shadow. Silver behenate ($d_{(100)}$ = 58.38 Å) is used as standard to calibrate the q-scale. Data are normalized by the transmission and calibrated to the SAXS signal of $H_2O$ at large q-values (q= 0.0163 cm$^{-1}$) in order to obtain an absolute intensity scale. The water signal is measured by subtracting the signal of the empty capillary from the signal of a water-filled capillary. The signal of (water+capillary) is used as background for the samples and it is subtracted after integration of the 2D data. For a consistent background subtraction, all data are recorded on the same capillary at the exact same spot throughout the experiment.

**Table S 1 - Detailed information concerning the SLC16:0 lipid solution during in situ SAXS experiment.**

| HCl solution molarity used to reduce pH | Rate of addition of HCl solution / μL/s | Initial pH of the lipid solution | Final pH of the lipid solution | Time to reach pH 3 in the lipid solution / min | Initial volume of lipid solution / mL | Concentration of lipid (SLC16:0) in solution / wt% | Total added volume of acid solution to lipid solution / μL | Dilution factor of lipid after adding HCl solution | Final NaCl concentration in lipid solution after adding HCl solution / mM |
|---|---|---|---|---|---|---|---|---|---|
| 0.1 | 0.136 | 9.6 | 2.3 | 50 | 2 | 0.5 | 415 | 17 % | 17.2 |
| 0.5 | 0.136 | 9.6 | 2.3 | 13 | 2 | 0.5 | 32 | 1.6 % | 7.9 |

Although not in a flow-through mode, the above configuration is employed to measure the signal of the SLC16:0 hydrogels prepared at C= 1 wt% and 2.5 wt% at pH 5.

The SAXS study of SLC16:0 hydrogels containing NaCl are performed in 2 mm quartz capillaries analysed at the B21 beamline of DIAMOND synchrotron (Oxford, UK). The energy of the beam is set at 12.4 KeV and a sample-to-detector distance of 4.014 m. The signal of the detector, used to record the data, is integrated azimuthally using the beamline software and in order to obtain the *I(q) vs. q* spectrum. Silver behenate ($d_{(100)}$ = 58.38 Å) is used as standard to calibrate the q-scale. Data are normalized by the transmission and calibrated to the SAXS signal of $H_2O$ at large q-values (q= 0.016 cm$^{-1}$) in order to obtain an absolute intensity scale. The background (water+capillary) is recorded using a new capillary and it is subtracted from the integrated data.

*Analysis of the SAXS data.* The pH-resolved *in-situ* SAXS data were analyzed using a model-dependent and model-independent approach. The low-q region below q< 0.03 Å$^{-1}$ is analyzed using a classical analysis of the slope of the I(q) data on a log-log scale. The data are fitted using a linear function which provides a value generally related to a specific morphology (e.g., spherical, cylindrical, lamellar),[1] or describing the presence of fractal objects.[2] In this work, we divide the region in a low-q, between -∞ < q / Å$^{-1}$ < ~0.03, and mid-q, between ~0.03 < q / Å$^{-1}$ < ~0.1, to account for the different scattering behaviour observed in this range. The q-region above 0.03 Å$^{-1}$ is also fitted with a model-dependent function, the general expression of which is I(q) ∝ P(q) S(q), where P(q) is the form factor of the scattering object and S(q) being the structure factor correlating objects in space. At infinite dilution, S(q) is unity and I(q) becomes proportional to P(q) only. Here, we make the approximation that the volume fraction used here (0.5 w%) is low enough to consider the approximation S(q)~ 1 valid at high q values above 0.03 Å$^{-1}$.

*Fit of SAXS data at q> 0.03 Å$^{-1}$: micelle model.* In a recent series of publications,[3,4] it was shown that the SAXS profile of acidic SLC18:1 and SLC18:0 sophorolipids at pH 11 and 5 wt% is practically the same. It was also shown that a core-shell ellipsoid of revolution form factor model is the most appropriate to fit the data in this q-region, thus identifying the presence of micellar aggregates. This result was used here to fit all samples at pH between 10.4 and 6.3. We used the same model developed in the SasView 3.1.2 software (CoreShellEllipsoidXT),[5] the general equation of which is

$$I(q) = \frac{scale}{V}\left(\rho - \rho_{solv}\right)^2 P(q)\, S(q) + bkg \qquad\qquad \text{Eq. S1}$$

where, $scale$ is the volume fraction, $V$ is the volume of the scatterer, $\rho$ is the Scattering Length Density (SLD) of the object, $\rho_{solv}$ is the SLD of the solvent, $P(q)$ is the form factor of the object, $bkg$ is a constant accounting for the background level and $S(q)$ is the structure factor, which is hypothesized as unity in the analyzed range of q-values. The analytical expression of the $P(q)$ for a core-shell ellipsoid of revolution model implemented in the software is provided in ref. [5], while Figure S 3a shows the geometrical model, where T is the equatorial shell thickness, $x$T is the polar shell thickness, R, the equatorial core radius, $y$R, the polar core radius. The model implies evaluating the $\rho_c$, $\rho_s$, $\rho_{solv}$, the SLDs of, respectively, the micellar core, shell and solvent. The model also considers a non-homogeneous core and shell, for we define the aspect ratio of the core radius and shell thickness, respectively being $x$ and $y$. Some fitting parameters can be simply estimated from the sample composition and they have been fixed throughout the fitting process. Among them, the volume fraction of scattering objects of 0.5 w% is kept constant. The core and solvent SLD have been fixed and calculated using Eq. S2 through the SLD calculator tool available in the SasView 3.1.2 software:

$$\rho = \frac{\sum_{i}^{j} Z_i r_e}{v_M} \qquad\qquad \text{Eq. S2}$$

where $Z_i$ is the atomic number of the $i^{th}$ of $j$ atoms in a molecule of molecular volume $v_M$, $r_e$ is the classical electron radius or Thomson scattering length ($2.8179 \times 10^{-15}$ m). $\rho_c$ is set to $8.4 \times 10^{-6}$ Å$^{-2}$, a typical value for a hydrocarbon chain in sophorolipids,[6] and $\rho_{solv}$ is set to $9.4 \times 10^{-6}$ Å$^{-2}$, a known value for water. The shell SLD, $\rho_s$, accounting for the carbohydrate moieties, water and counterions, is always a variable parameter; nevertheless, a reasonable assumption, which also constitutes an internal standard to check the fit quality, is that $\rho_s$ should be contained between the hydrated and dehydrated sophorose, that is between 10.0 and $14.0 \times 10^{-6}$ Å$^{-2}$. This range is estimated with the SLD calculator available in the 3.1.2 version of SasView software. For the low range, one can use formic acid, as a model carboxylic acid ($CH_2O_2$, d= 1.2 g/cm$^3$, SLD= $10.7 \times 10^{-6}$ Å$^{-2}$). For the high range, one can use glucose as a model for sophorose ($C_6H_{12}O_6$, d= 1.6 g/cm$^3$, SLD= $14.5 \times 10^{-6}$ Å$^{-2}$).[6] The overall quality of the fit is followed by the classical $\chi^2$ evolution test. The estimated error on the fitted parameters is about $\pm 10\%$.

*Fitting strategy.* For the pH-dependent experiments, we start from the best fits obtained at pH 9.6 and we use the batch fitting mode developed in the SasView 3.1.2 software to fit all data above q> 0.03 Å$^{-1}$ employing the same form factor. To find consistency along the batch fitting, we only fitted twenty curves at a time from pH 9.6 to pH 5.5, where we use the best fit values of the 20$^{th}$ fit as input parameters to fit the 21$^{th}$ - 40$^{th}$ curves and so on. In the first fitting session, R, T, $x$, $y$ and $\rho_s$ are all set as variable parameters. The results of this initial fitting process are shown in Figure S3a-c in the Supporting Information. However, it seemed that the core aspect ratio, $x$, had little influence on the fit. For this reason, we estimated an optimal value of $x$= 3.5 from Figure S3a and kept $x$ constant throughout the a second fitting session, where only four parameters are eventually kept as free variables. The fitted values given in Figure S3 show the results from both $x$= 3.5 and $x$= variable. Comparison between the fit results shows the same overall trend. $\chi^2$ evolution and value of $\rho_s$ are always used as controls to follow the quality of the fit.

*Rheology.* Rheology experiments are carried out using a MCR 302 rheometer (Anton Paar, Graz, Austria) equipped with a Peltier temperature system which allows accurate control of the temperature by the stainless steel lower plate and with a solvent trap.

- *Oscillatory rheology.* These experiments conducted at the controlled T= 20°C are performed using a stainless steel sandblasted upper plate (diameter 25 mm). The gap (0.5 mm) and the normal force (NF= 0 N) are controlled during the experiments. Samples are vortexed for few seconds before loading on the rheometer. After loading, samples are allowed to stand at rest for thirty minutes before analysis. Standard angular frequency sweep analysis (100 – 0.06 rad·s$^{-1}$) was performed using a shear strain within the linear viscoelastic regime (LVER), $\gamma$= 0.1%, verified by a dynamic strain sweep conducted at an angular frequency ($\omega$= 6.28 rad·s$^{-1}$) by varying the shear strain ($\gamma$) from 0.001 to 100 %.

- *Step-strain experiments: study of thixotropy.* These experiments were performed between 1 h and 3 h from sample loading and consisted of applying a destruct and recovery cycle series. Unless otherwise specified, the first cycle consists in 30 s with $\gamma$= 100% and 10 min recovery at $\gamma$= 0.1%. All other cycles involve 5 min with $\gamma$= 100% and 10 min recovery at $\gamma$= 0.1%.

*Cryogenic Transmission Electron Microscopy (Cryo-TEM).* These experiments were carried out on an FEI Tecnai 120 twin microscope operating at 120 kV equipped with a Gatan

Orius CCD numeric camera. The sample holder was a Gatan Cryoholder (Gatan 626DH, Gatan). Digital Micrograph software was used for image acquisition. Cryofixation was done on a homemade cryofixation device. The solutions were deposited on a glow-discharged holey carbon coated TEM copper grid (Quantifoil R2/2, Germany). Excess solution was removed and the grid was immediately plunged into liquid ethane at −180 °C before transferring them into liquid nitrogen. All grids were kept at liquid nitrogen temperature throughout all experimentation. TEM images have been analyzed with Fiji software.[7]

*Differential Scanning Calorimetry (DSC)*: DSC is performed using a DSC Q20 apparatus from TA Instruments equipped with the Advantage for Q Series Version acquisition software (v5.4.0). Acquisition is performed on a dry powder sample (~ 8 mg) sealed in a classical aluminium cup and using an immediate sequence of heating (from 10°C to 90°C) and cooling (from 90°C to 10°C) ramps both at a rate of 1°C/min.

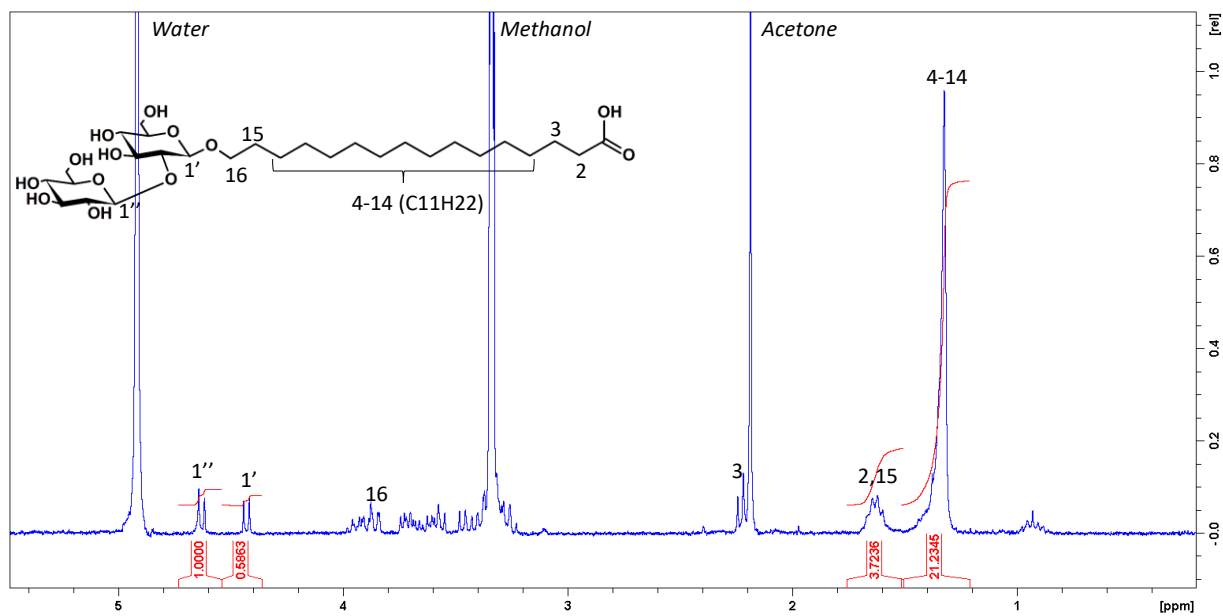

**Figure S 1 – ¹H NMR spectrum of SLC16:0 sophorolipids (solvent MeOD-d4). Integration of the 4-14 peak at 1.325 ppm reflects the C16:0 backbone.**

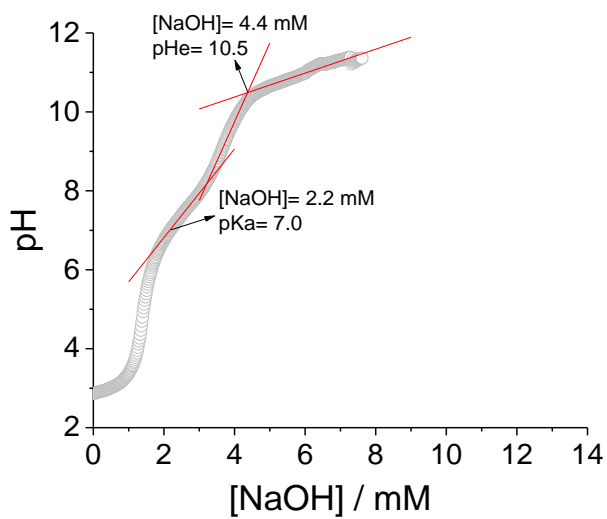

**Figure S 2 – Titration curve of SLC16:0 at C= 0.5 wt% (volume: 0.5 mL) from pH 3 to pH 11. Sample is prepared by diluting five times a hydrogel initially prepared at C= 2.5 wt% and pH 3. The hydrogel is prepared according to the pH jump method described in the materials and method section. Titration is performed with a 0.5 M NaOH solution added to the lipid solution at a rate of 0.2 µL/min using an automated push-syringe. pH is measured using a Mettler Toledo microelectrode and a Hanna Instrument pH-meter using the provided acquisition software. pH is sampled every 5 s.**

**Model-dependent analysis of pH-resolved *in situ* SAXS experiments**

On the basis of previous SAXS data recorded on similar microbial glycolipid amphiphiles,[3,4,8,9] we employ the ellipsoid (prolate) of revolution form factor model function developed in the SasView (3.1.2 version) software package,[5] characterized by chemically-different hydrophobic core and hydrophilic shell regions of uneven thickness along with the equatorial and polar directions (Figure S 3a). The same model was described in detail and applied to the successful fit of SAXS data recorded for other microbial glycolipids at basic pH.[3,4,9] This model is relatively complex (nine independent parameters), compared to single-density sphere, cylindrical or core-shell sphere/cylindrical models. If one considers the chemical nature of SLC16:0 in particular, and microbial glycolipids in general, the use of a complex model is risky but not outrageous. Compared to classical head-tail surfactants, which can easily modeled with a double head(shell)-tail(core) density distribution,[10] glycolipid biosurfactants are characterized by a bulky sugar region, which sets at the water interface, an aliphatic region, occupying the center of the micelle, and a mixture of carboxylate/carboxylic acid groups, of which the location is not intuitive.[6] In this case, a core-shell model, where the shell is composed of at least glucose and most-likely the carboxylic acid group and the core is occupied by at least the aliphatic region, is at least expected.[3,6,11] However, it is not unreasonable to suppose that the hydrophilic shell region is most-likely not homogeneous, if one considers the coexistence of both carboxylate and carboxylic acid groups in the same micellar aggregate. They have different hydration properties and induce different spontaneous curvature, their ratio varies with pH and in this sense they could occupy different regions of the micelle, thus inducing an asymmetry in the homogeneity of the hydrophilic shell.

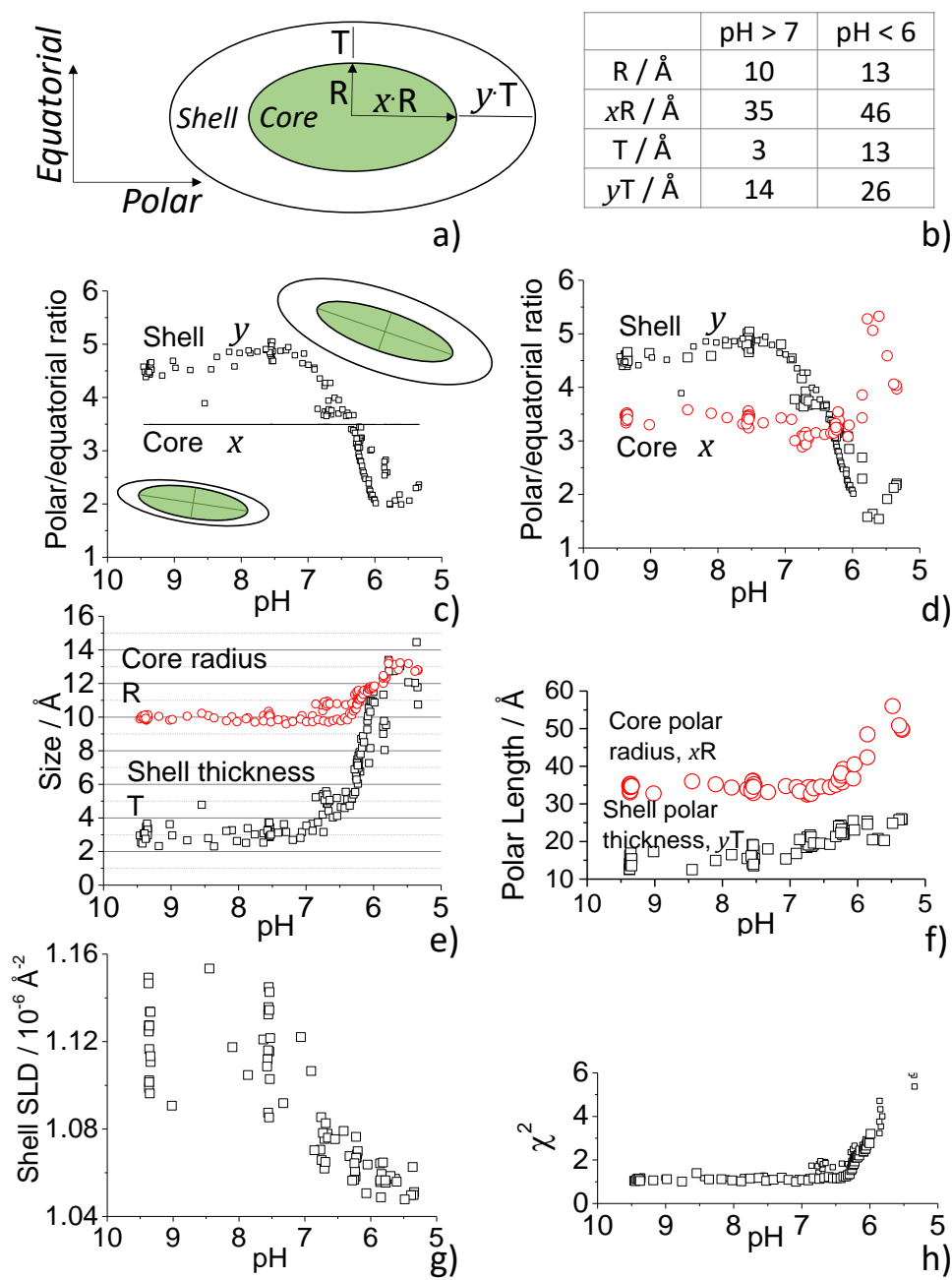

**Figure S 3** - Evolutions of the structural parameters obtained from the fit of the pH-resolved *in situ* SAXS data collected on the SLC16:0 system acidified with a **0.5 M HCl solution** and presented in Figure 2 of the main text. a) Core-shell (prolate) ellipsoid of revolution model form factor.[5] b) Typical values of the equatorial and polar core radii (R, *x*R) and shell thickness (T, *y*T) at acidic and basic pH. c,d) pH-dependent evolution of the polar-to-equatorial shell (*y*) and core (*x*) ratio. Please note that in c), (*x*) is voluntarily fixed at the constant value of 3.5 (straight line), while in d) (*x*) is fitted as a free parameter. e-g) pH-dependent evolution of e) the core radius (R) and shell thickness (T), f) core and shell polar lengths (respectively, *x*R and *y*T), g) scattering length density (SLD) of the shell. g) Evolution of $\chi^2$ for both *x* = 3.5 and *x*= variable.

**Notes: 1)** The core and shell dimensions in the core-shell ellipsoid drawings in c) are scaled to their fitted dimensions. **2)** The plotted values of shell *y*, size (R, T), polar length (*x*R, *y*T), SLD and $\chi^2$ in d-h) are

obtained from two sets of fits, where $x$= 3.5 and $x$= variable. 3) The core and shell polar length, $x$R and $y$T, in f) are calculated as the product of the polar/equatorial ratio and size, respectively given in d) and e).

Despite its assumed complexity, the model used in this work is still the simplest micellar model that allows a satisfactory fit of the SAXS data. In fact, any simpler one, including the more common core-shell ellipsoid with a homogeneous shell thickness[12] or core-shell sphere with homogeneous shell[13] and enhanced polydispersity, fail. The model shown in Figure S 3a considers that the shell region is not homogeneous in the polar and equatorial directions. The corresponding physical interpretation should not go beyond the hypothesis that the composition of the hydrophilic shell around the core is in fact poorly defined, and it has a huge impact on the amplitude and broadness of the oscillation of the form factor above q= 0.1 Å$^{-1}$. Despite its apparent complexity, quantified in nine independent fitting parameters, one can reduce them to only five critical ones.

All fits have the following fixed parameters, as discussed previously:[14] background= 0.0005 cm$^{-1}$ and scale= 0.005, the latter corresponding to the experimental volume fraction of 0.5 wt% considering the absolute scale of our SAXS data. The scattering length densities (SLD) of the hydrophobic core and solvent are, respectively, $\rho_c$= 8.4·10$^{-6}$ Å$^{-2}$ and $\rho_{solv}$= 9.4·10$^{-6}$ Å$^{-2}$, both calculated for hydrocarbons and water. Finally, $x$ is set both as a variable (Figure S 3d) and at $x$= 3.5 (straight line in Figure S 3c), as better detailed below.

All independent variables (R, T, $x$, $y$ and shell SLD) are initially refined by hand according to the values obtained at basic pH for other glycolipids with similar behaviour,[14] and they are eventually fitted. A quick analysis of the pH-dependent evolution of the core aspect ratio, $x$, in Figure S 3d shows that this parameter is practically constant at the average value of 3.5 and it strongly fluctuates at pH below 6. To reduce the number of variables, we then tested a second fitting approach with $x$= 3.5 (Figure S 3c) and we find no significant differences in terms of the pH-dependent evolution of all parameters. For this reason, the graphs in Figure S 3d-h report the data obtained with both $x$= 3.5 and $x$= variable. This approach has the double advantage of showing the robustness of the fitting strategy and reducing the fitting complexity to only four variables (R, T, $y$ and shell SLD) when $x$= 3.5.

The typical fits are superimposed (red curves) on the corresponding, selected, SAXS profiles at pH 9.38 and 6.54 (Figure 2b in the main text). We experience a good match between the fit and the experimental data, as also shown by the corresponding pH-dependent evolution of $\chi^2$ (Figure S 3h) and contained between 2 and 6. The increase of $\chi^2$ below pH 6 is significant and it indicates that the ellipsoid model becomes less and less reliable to fit the data, due to the coexistence between different structures affecting the same region of q-values. The model-dependent approach should then used only until pH 5.5. Last, the fit is performed only between q= 0.4 Å$^{-1}$ and q= 0.03 Å$^{-1}$, below which the model-independent slope analysis is more safely employed, due to the presence of larger, ill-defined, structures. This is not an obstacle to our approach, because the micellar environment of SLC16:0 (typical size below about 30 Å) is rather probed in the q-region below q= 0.03 Å$^{-1}$ (typical size domain of 200 Å), and in particular in the oscillation profile of the form factor. Even if one cannot exclude that keeping away the region below q= 0.03 Å$^{-1}$ in the fitting process may introduce an error in the absolute values of the micellar structural parameters reported in Figure S 3d-g, the overall trend of each parameter as a function of pH is not affected. Values given in Figure S 3d-g should be considered within an error of ±10%.

According to the fit, at pH> 7 the micelle can be described as a prolate ellipsoid of revolution (Figure S 3a) with an equivalent equatorial core radius, R, and shell thickness, T, of respectively 10 Å and 3 Å, and an overall polar dimension ($x$R+$y$T) of 49 Å (Figure S 3b,f). The shell aspect ratio, $y$, settles at about 4.5 (Figure S 3d). These structural parameters are constant until pH ~7. Between pH 7 and 6, the equatorial thickness, T, increases by a factor 3, while the core radius, R, increases by a factor 1.3 (Figure S 3e); in the meanwhile, the overall polar direction ($x$R+$y$T) increases by a factor 1.5, whereas the polar shell thickness, $y$T, is only twice as thick as T (Figure S 3f).

The graphical evolution of the equatorial and polar core and shell radius and thickness parameters, scaled to the experimental data, are represented in the core-shell ellipsoid drawings shown in Figure S 3c. An increased homogenization of the equatorial and polar shell thickness at lower pH (decreasing $y$; Figure S 3c,d) seems to be a common feature found for acidic sophorolipids.[3] Similar to C18:1 sophorolipids, the pH-evolution of the oscillation of the form factor with pH indicates a local redistribution of the lipids when they pass from their ionic to acidic forms. The constancy in terms of the micellar structural parameters until pH 7 could be strongly related to the pKa of SLC16:0, which is found to be 7.0 by titration (Figure S 2). In other words, when the amount of negative charges reduces by half (COO$^-$ ≈ COOH), the electrostatic repulsion between carboxylate groups becomes less important and the micellar curvature decreases, thus inducing a more homogeneous distribution of the carboxylic acids, which experience less electrostatic repulsions between pH 7 and pH 6. It then becomes more intuitive to explain the unusual non-homogeneous evolution of the shell thickness, compared to classical surfactants. The larger shell thickness in the polar direction compared to the equatorial one above pH 7 could be explained by a local accumulation of COO⁻ groups, better withstanding higher curvatures. On the contrary, the COOH group will rather settle in the equatorial, less curved, direction below pH 7. Finally, the shell scattering length density, SLD, seems to decrease with pH, suggesting a tendency of the shell towards hydration. However, any conclusion based on the values of SLD should be taken with caution, due to its strong fluctuation above pH 7.

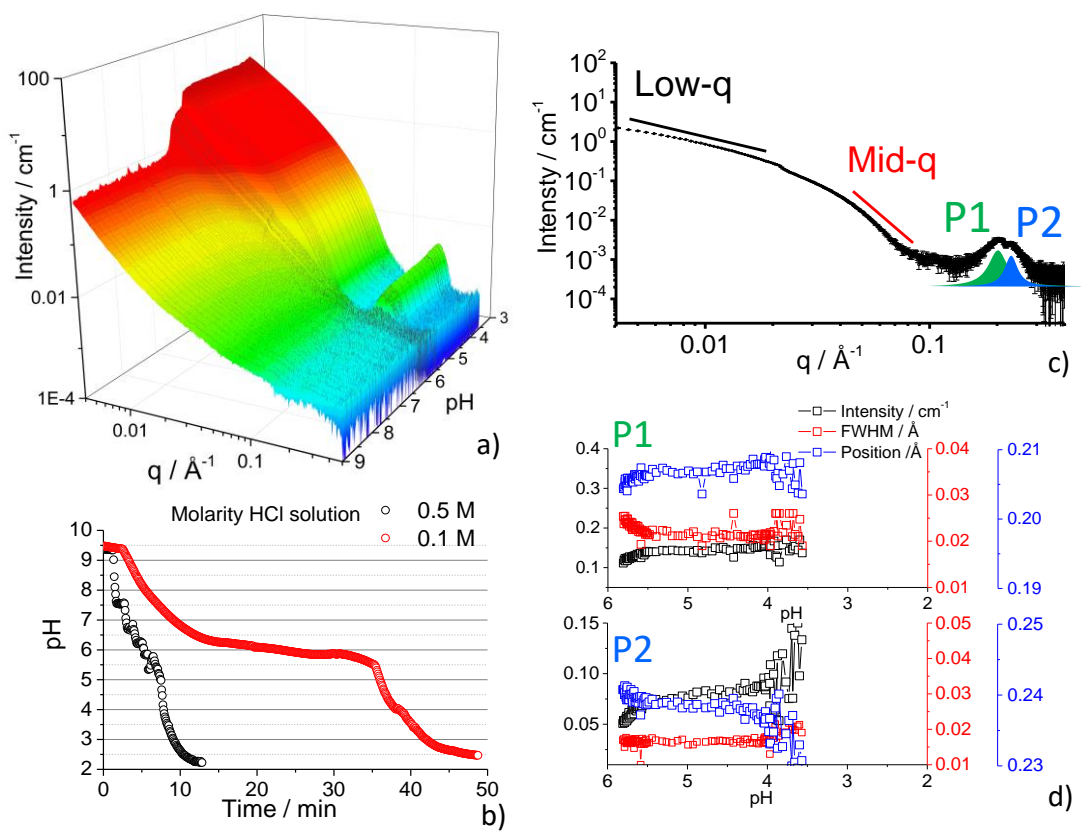

**Figure S 4 - a)** pH-dependent *in situ* SAXS profiles recorded for SLC16:0 solution (V= 2 mL) at T= 23°C and C= 0.5 wt% in water. pH is varied from basic to acidic by adding micromolar amounts of <u>HCl solution at 0.1 M</u> (rate: 0.136 µL/s, total volume added, 415 µL, dilution factor= 17%). **b)** Evolution of the pH with time when 0.1 M and 0.5 M HCl solutions are employed (rate: 0.136 µL/s) to acidify a SLC16:0 solution (V= 2 mL) at C= 0.5 wt%. **c)** Typical SAXS profiles extracted at pH< 5 from a). Continuous black and red lines indicate the q-domains of linear fits in log-log scale while P1 and P2 schematically show the Lorentzian peaks used to fit the experimental diffraction peak above 0.1 Å⁻¹. **d)** pH-resolved evolution of the full structural parameters of P1 and P2.

*Comment on the coexistence of two diffraction peaks in the in situ SAXS data at acidic pH.*

The presence of two concomitant peaks instead of one is puzzling, especially because their respective positions (q(P1):q(P2) ~0.83), which do fall in classical lamellar (1:2) or hexagonal (1:1.71) orders, full width at half maximum (FWHM) and intensity independently evolve with both pH and conditions of synthesis. To illustrate this fact, The figure below plots the intensity, FWHM and position of P1 and P2 from pH 5.5 to pH 2 for the system acidified with a 0.5 M HCl solution, while Figure S 4c,d plots the same parameters for a system acidified with a 0.1 M HCl solution. In the former, P1 becomes sharper and more intense while P2 also becomes sharper but its intensity does not vary in the pH range explored. Under these conditions, P1 overwhelms P2 when pH is decreased. In the second case scenario, when acidification occurs at a slower rate (use of a 0.1 M HCl solution, Figure S 4b), both P1 and P2 seem to be equivalent in terms of position, intensity and FWHM. The origin of P1 and P2 is discussed in the main text and it is most likely associated to the presence of two polymorphs of SLC16:0.

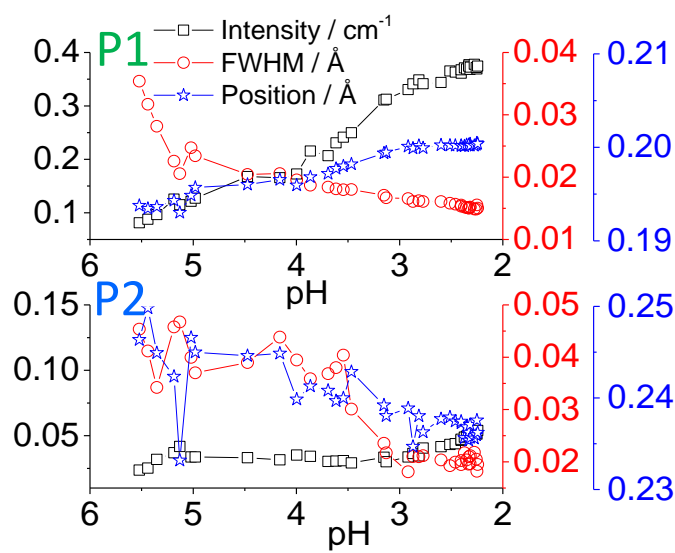

**pH-resolved evolution of the full structural parameters of P1 and P2 corresponding to the system in Figure 2 and Figure 3 in the main text**

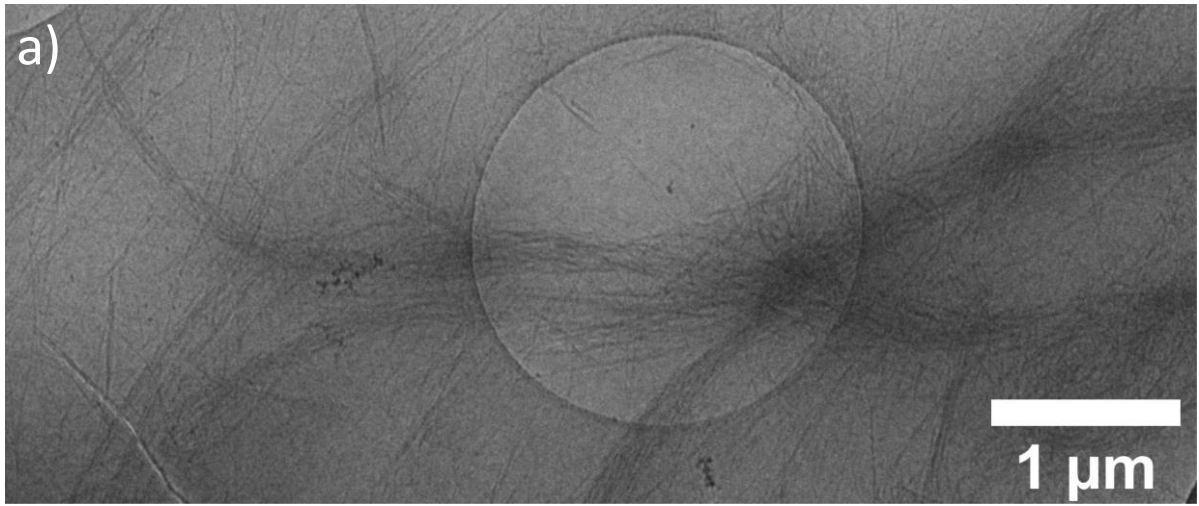

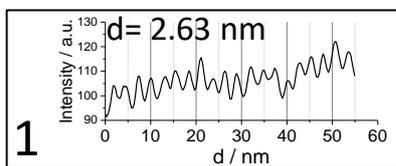

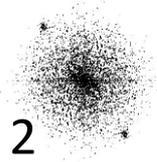

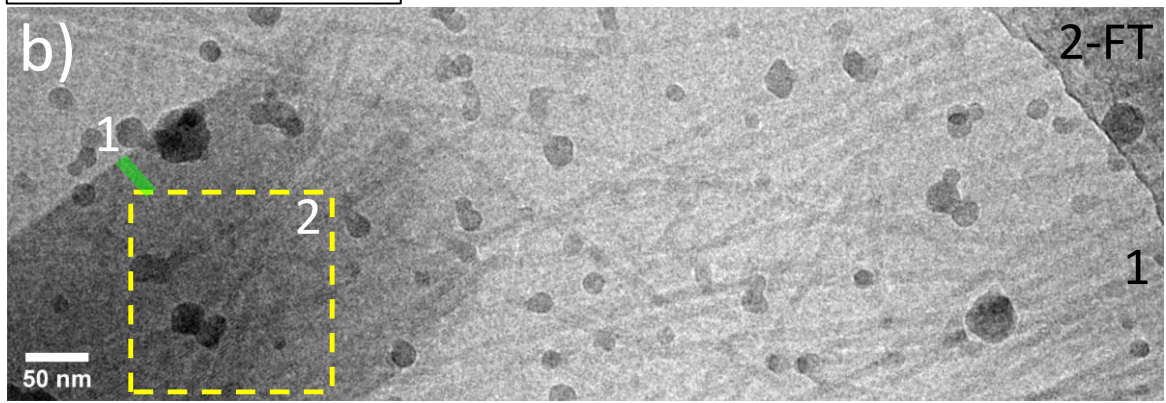

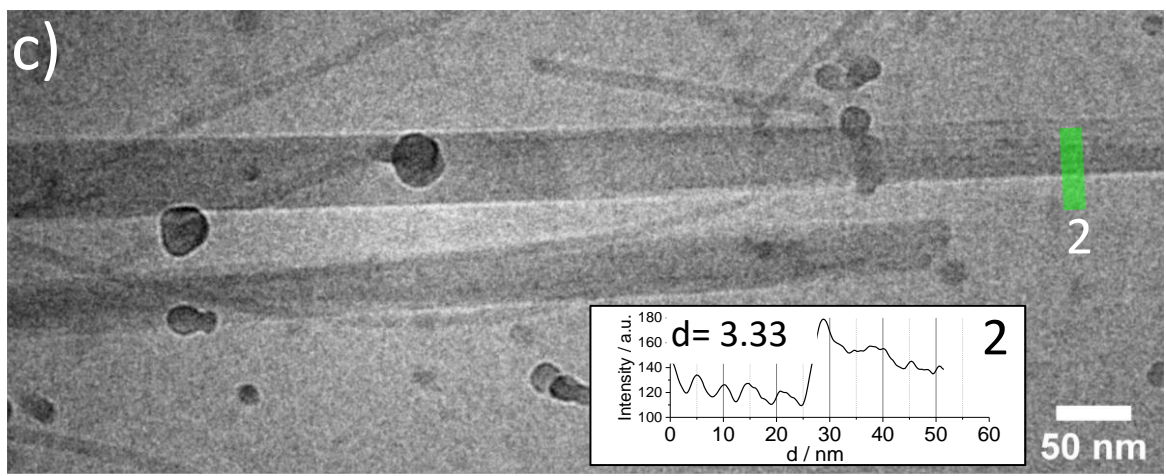

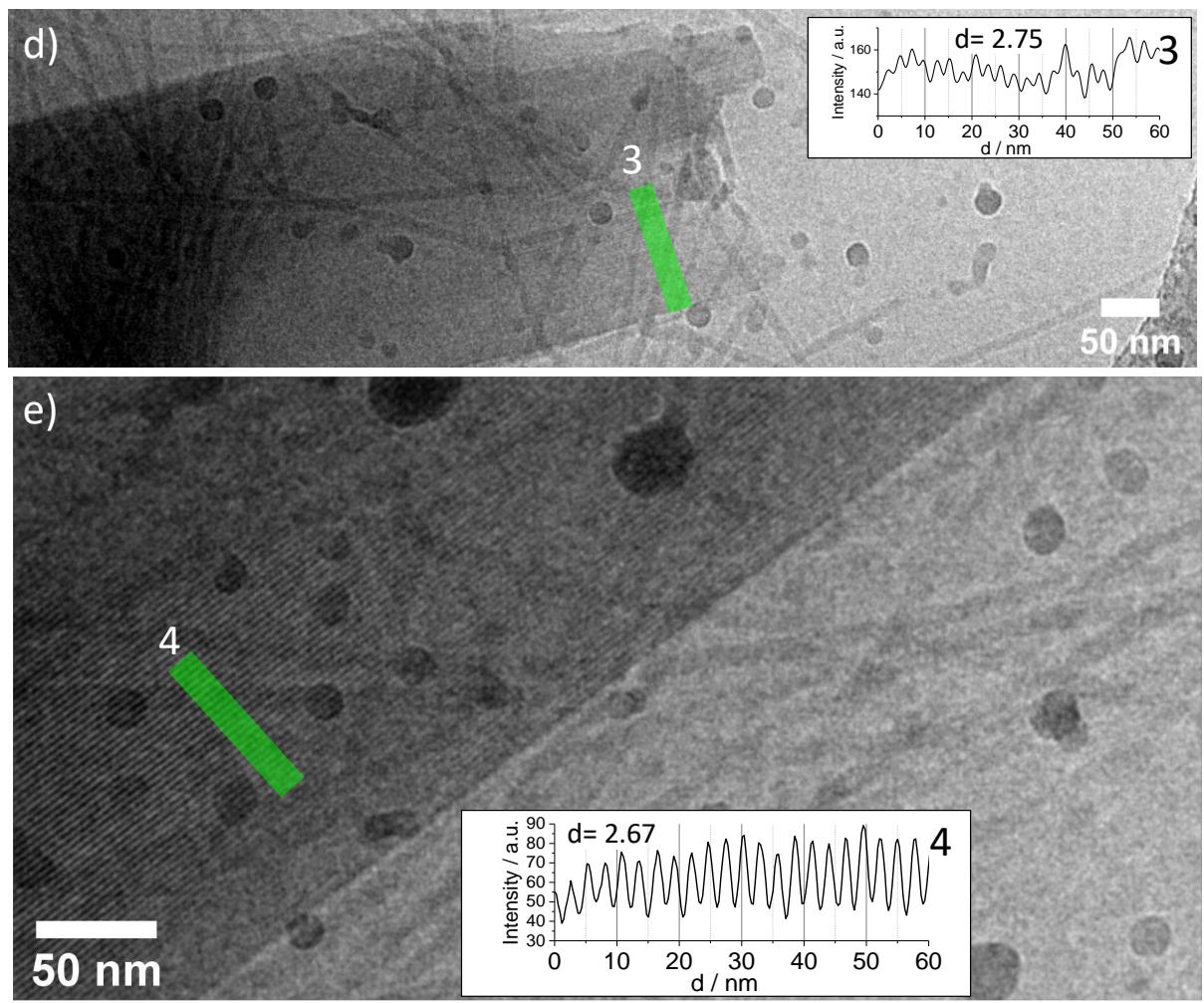

**Figure S 5 – Complementary cryo-TEM images recorded on a SLC16:0 sample at C= 0.25 wt% and pH 3. Sample is diluted ten times from a hydrogel prepared at C= 2.5 wt%, pH= 3 using the pH jump approach described in the materials and method section.**

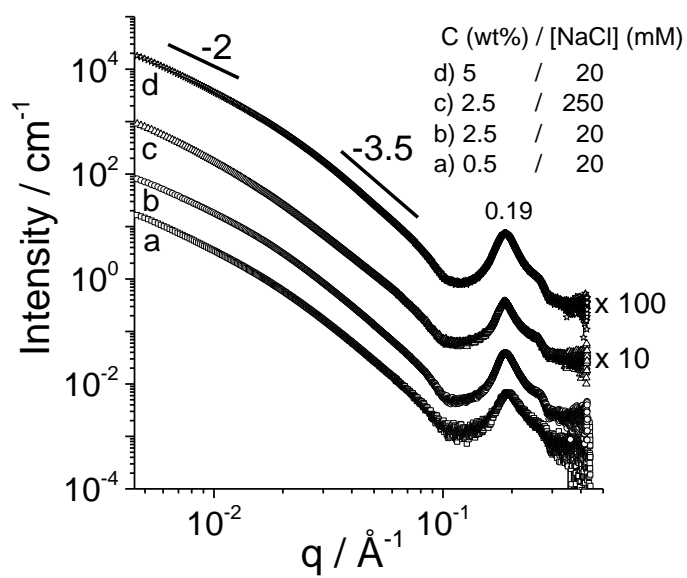

**Figure S 6 – SAXS profiles recorded on SLC16:0 samples prepared at concentration values between C= 0.5 wt% and 5 wt% at pH 5 and for various ionic strengths. Profiles c and d are respectively multiplied by a factor of 10 and 100 for better reading. These experiments were recorded at the B21 beamline of Diamond synchrotron, Oxford, UK.**

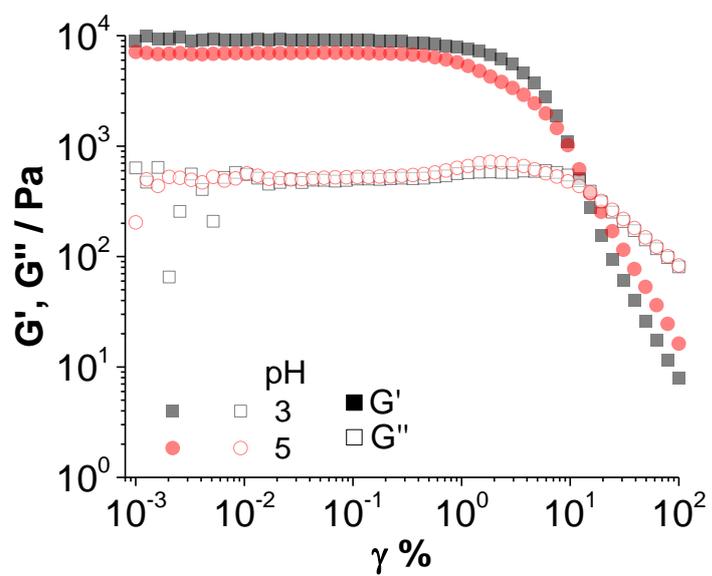

**Figure S 7 – Typical strain sweep experiments recorded on a SLC16:0 sample prepared at C= 2.5 wt% and pH= 3 and pH= 5 (T= 20°C) using a plate-plate geometry and ω= 6.28 rad.s⁻¹**

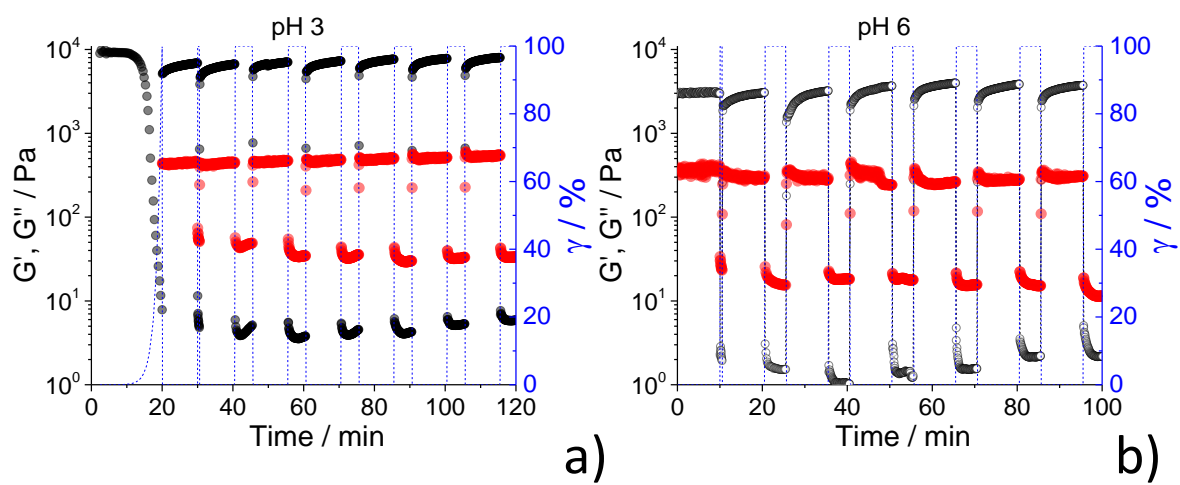

**Figure S 8 – Study of the recovery properties of SLC16:0 hydrogels (C= 2.5 wt%, T= 20°C, ω= 6.28 rad/s) at a) pH= 3 and b) pH= 6.**

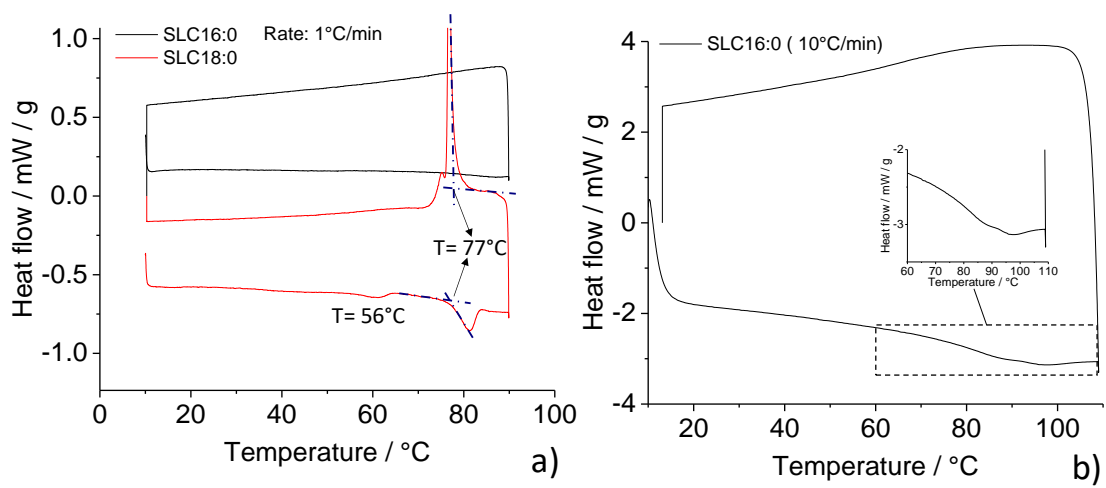

**Figure S 9 – a) Differential Scanning Calorimetry (DSC) thermograms of the deacetylated acidic sophorolipid SLC16:0 and SLC18:0 dry powders. The samples are equilibrated at T= 10°C, followed by a ramp of 1°C/min until T= 90°C and then back to T= 10°C using the same cooling rate. b) DSC thermogram of SLC16:0 powder between 10°C and 110°C at a heating and cooling rates of 10°C/min.**